\documentclass[prd,showpacs,nofootinbib,preprint]{revtex4}
\usepackage{amsmath}
\usepackage{amsfonts}
\usepackage{graphicx}
\graphicspath{{Figures/}}
\DeclareGraphicsExtensions{.pdf,.png,.jpg,.eps}

\def\be{\begin{equation}}
\def\ee{\end{equation}}

\numberwithin{equation}{section}

\begin{document}

\title{Leakage of gravitational waves into an extra dimension in the DGP model}

\author{M. Khlopunov}
\email{khlopunov.mi14@physics.msu.ru}
\affiliation{Faculty of Physics, Lomonosov Moscow State University, Moscow, 119899, Russia\\
Institute of Theoretical and Mathematical Physics, Lomonosov Moscow State University, Moscow, 119991, Russia}

\author{D.V. Gal'tsov}
\email{galtsov@phys.msu.ru}
\affiliation{Faculty of Physics, Lomonosov Moscow State University, Moscow, 119899, Russia}

\begin{abstract}
In the DGP model, the graviton is unstable, which leads to a modification of gravity at cosmological distances. In particular, this leads to the leakage of gravitational waves from the brane into an extra dimension at large distances from the source. However, the calculation of the gravitational wave leakage intensity is a non-trivial task due to the violation of the Huygens principle in the five-dimensional bulk of the DGP setup. The odd dimension of the bulk makes it difficult to extract the radiated part of the field. In this paper, we consider a simplified problem of scalar radiation from a point charge localized on a brane in the framework of the scalar field analog of the DGP model. In this model, the scalar field on the brane can be represented as a continuous spectrum of Kaluza-Klein massive modes. To extract the emitted part of such a field, we generalize the Rohrlich-Teitelboim approach to radiation to the case of a massive four-dimensional field, using its connections to massless fields in four and five dimensions. In the case of a charge moving along a circular trajectory, we obtain the dependence of the radiation energy flux through a 2-sphere localized on the brane on the sphere radius, which provides the intensity of leakage of scalar radiation from the brane. Consistent with the infrared transparency of the bulk, the leakage intensity is found to be higher for low frequency signals. We are also analyzing the possibility of detecting this leak by current and future gravitational-wave observatories.
\end{abstract}

\maketitle

\section{Introduction}

Extra spacetime dimensions are an essential ingredient of a number of modern theories. Thus, the string theory, being the main model of quantum gravity, predicts the existence of extra dimensions \cite{Green:1987sp}. Extra dimensions are inherently present in the holography \cite{Arefeva:2014kyw,Cardoso:2013vpa}. Also,  over the past twenty years a number of phenomenological theories   with extra dimensions has been constructed to solve the hierarchy problem \cite{Arkani-Hamed:1998jmv,Randall:1999ee,Randall:1999vf} and puzzles of cosmology \cite{Dvali:2000hr,Deffayet:2000uy,Deffayet:2001pu,Deffayet:2002sp} (for review see, e.g., \cite{Rubakov:2001kp,Maartens:2010ar,Cheng:2010pt}). Recently, actively developing gravitational-wave astronomy opens up new ways of experimental study of extra dimensions (see, e.g., \cite{Yu:2019jlb,Ezquiaga:2018btd}). 
Extra dimensions can manifest as additional polarizations of gravitational waves \cite{Andriot2017,Khlopunov:2022ubp}, tower of massive high-frequency Kaluza-Klein modes of the gravitational-wave signals \cite{Andriot2017,Andriot:2019hay,Andriot:2021gwv,Barvinsky:2003jf}, additional contributions to the source of gravitational field \cite{Shiromizu:1999wj,Maeda:2003ar,Garcia-Aspeitia:2013jea,Kinoshita:2005nx}, differences in propagation of gravitational-wave and counterpart electromagnetic signals \cite {Yu:2016tar,Visinelli:2017bny}, as well as in quasi-normal modes \cite{Chakraborty:2017qve,Mishra:2021waw} and tidal deformations \cite{Chakravarti:2018vlt,Cardoso:2019vof,Chakravarti:2019aup} of black holes and neutron stars. Photographs of black hole shadows \cite{Vagnozzi:2019apd,Banerjee:2019nnj,Neves:2020doc,Banerjee:2022jog} are another promising tool for exploring extra dimensions.

In this paper, we consider the DGP braneworld model \cite{Dvali:2000hr,Dvali:2000xg} with one extra spacelike dimension. In this model, an effective four-dimensional graviton on the brane is a metastable resonance with a finite lifetime \cite{Gabadadze:2003ck,Gabadadze:2004dq} (see, also, \cite{Dvali:2000rv,Gregory:2000jc,Csaki:2000pp,Csaki:2000ei}).
As a result, in the DGP model, the laws of gravity change at large distances from the source of the gravitational field. In particular, this leads to the fact that one of the branches of the DGP model contains a cosmological solution with a late-time self-accelerating brane \cite{Deffayet:2000uy,Deffayet:2001pu,Deffayet:2002sp}. 
Although a ghost degree of freedom \cite{Gorbunov:2005zk,Charmousis:2006pn} was found in the corresponding branch of the DGP model, which makes it problematic to solve the cosmological constant problem within this model, the second branch of the DGP model, called normal, remains an interesting diffeomorphism-invariant model of the metastable graviton. In particular, this should lead to a faster attenuation of the amplitude of gravitational waves with cosmological scale distances from the source due to leakage of gravitational waves into an extra dimension \cite{Deffayet:2007kf}. Thus, using joint observations of gravitational-wave and electromagnetic signals from mergers involving neutron stars, it becomes possible to limit the parameters of the DGP model \cite{Deffayet:2007kf,Pardo:2018ipy,Corman:2020pyr,Corman:2021avn,Hernandez:2021qfp}. The aim of this paper is to calculate the intensity of the leakage of gravitational waves into the extra dimension within the simple framework of the scalar field analog of the DGP model interacting with the point charge localised on the brane. We, also, discuss the possibility of the experimental detection of this leakage by use of the current and future gravitational-wave observatories.

Here we would like to focus on the common feature of some extra-dimensional models (such as the DGP and Randall-Sundrum) combining even and odd-dimensional spacetimes. An important feature of odd dimensions is a violation of the Huygens principle, which was discussed in the classical works of Hadamard \cite{hadamard2014lectures}, Courant and Hilbert \cite{courant2008methods}, Ivanenko and Sokolov \cite{Ivanenko_book} and others. In the present context, the calculation of the intensity of the leakage of gravitational radiation into the bulk is hampered by the violation of the Huygens principle in the five-dimensional bulk of the DGP model. Behavior of massless Green's functions in the spacetime of even and odd dimensions is different. In even dimensions, the perturbation of the massless field from the instantaneous flash of the source, having reached the observation point in the time interval necessary for its propagation at the speed of light, stops instantly. On the contrary, in odd dimensions, an infinite tail signal decaying with time is preserved at the point of observation, corresponding to the propagation of massless fields in odd dimensions with all velocities up to the speed of light. Mathematically, this is due to the fact that the retarded Green's functions of massless fields in even dimensions are localized on the light cone, while in odd dimensions they are also localized inside it. At the same time, free massless fields propagate exactly at the speed of light in all dimensions. Thus, while the total retarded field of a localized source propagates at all velocities up to the speed of light in odd dimensions, its radiated part, which is the free field away from the source, must propagate exactly at the speed of light. This mismatch makes the extraction of the radiated part of the retarded field in odd dimensions a non-trivial task.
 
Because of this unusual behavior of the retarded massless fields in odd dimensions, most of the literature has studied only radiation problems in even dimensions \cite{Kosyakov1999,Cardoso:2002pa,Mironov:2006wi,Mironov:2007nk,Cardoso:2007uy,Kosyakov:2008wa} while the case of odd dimensions was discussed mainly in the context of the radiation reaction force \cite{Galtsov:2001iv,Kazinski:2002mp,Kazinski:2005gx,Yaremko2007,Shuryak:2011tt,Dai:2013cwa,Harte:2016fru} (see, also, \cite{Kosyakov:2007qc, Kosyakov:2018wek}). In odd dimensions, the radiated part of the field contains a non-local tail term similar to that found by DeWitt and Brehme in curved four-dimensional spacetime \cite{DeWitt:1960fc,Galtsov:2007zz,Barack:2018yvs}. However, while in the latter case the tail term is due to the scattering of the gravitational waves on the spacetime curvature and its computation is rather complicated, in odd-dimensional flat spacetime the tail term is known in closed analytic form. The tail term can be dealt with by use of the effective field theory approach to the problems of radiation \cite{Porto:2016pyg,Cardoso:2008gn,Birnholtz:2013ffa,Birnholtz:2015hua}. However, this approach does not provide us with information about the structure of the retarded field in the wave zone and the role of the tail term in the formation of radiation, because it is based on the calculations in momentum space insensitive to the dimensionality of the spacetime. As was shown recently, this problem can be dealt with in two ways: by Fourier transforming the retarded Green's functions over the temporal coordinate \cite{Chu:2021uea}, or by the modification of the definition of radiation  \cite{Galtsov:2020hhn,Galtsov:2021zpb,Khlopunov:2022ubp}. In this paper, we follow this second approach.

We use the Rohrlich-Teitelboim description of radiation \cite{Rohrlich1961,Teitelboim1970,rohrlich2007} (see, also, \cite{Kosyakov1992,Galtsov:2004uqu,Spirin2009,Galtsov:2010tny}), based on the Lorentz-invariant decomposition of the energy-momentum tensor of the retarded field, to extract the emitted part of the five-dimensional retarded DGP scalar field. As in the DGP model an effective four-dimensional field on the brane is decomposed into the continuum spectrum of massive Kaluza-Klein modes \cite{Gabadadze:2003ck,Dvali:2001gm,Dvali:2001gx,Gabadadze:2004jk} (see, also, \cite{Hinterbichler:2011tt,deRham:2014zqa}), we generalise the Rohrlich-Teitelboim approach to the case of massive four-dimensional field using its connections with the massless fields in four and five dimensions. As a result, in case of the non-relativistic charge moving along the circular trajectory on the brane, we find the dependence of the effective four-dimensional scalar radiation energy flux through the 2-sphere on the brane, covering the area of the charge motion, on its radius, which characterise the intensity of leakage of radiation from the brane. In accordance with the infrared transparency of the bulk \cite{Dvali:2000xg,Brown:2016gwv} in the DGP model, the intensity of the leakage of radiation is found to be higher for low-frequency signals. We also demonstrate that the obtained effective power of scalar radiation into the brane depends on the entire history of the charge's motion preceding the retarded time, rather than on its state at the retarded time moment, in contrast with the standard four-dimensional massless theory.

This paper is organised as follows. In section \ref{II}, we discuss the model under consideration, find the effective four-dimensional equation of motion for the field on the brane and the corresponding energy-momentum tensor localised on the brane, demonstrate on the simple two-dimensional example phenomenon of infrared transparency of the bulk in DGP model, and briefly recall the Rohrlich-Teitelboim approach to radiation. Section \ref{III} is devoted to the computation of the retarded Green's function of the field on the brane in DGP model. We illustrate the main steps of computation by simple lower-dimensional examples and, also, obtain the decomposition of the field on the brane into the continuum spectrum of massive four-dimensional canonical scalar fields. In section \ref{IV}, we find connections between the massive four-dimensional field and massless fields in four and five dimensions, which allow us to extract the emitted part of the former by use of the Rohrlich-Teitelboim approach, and compute it. Section \ref{V} is devoted to the computation of the effective four-dimensional scalar radiation energy flux through the 2-sphere on the brane. In a simple case of charge's circular motion, we explicitly compute the corresponding energy flux and analyse its properties. In section \ref{VI}, we discuss the obtained results.

\section{The Setup}\label{II}

Our goal is to determine the intensity of gravitational radiation leakage from the brane into the bulk in the DGP model. For simplicity, we neglect the tensor structure of the gravitational field and consider the scalar field analogue of the DGP model, which lives in the five-dimensional Minkowski space. We also neglect the fluctuations of the brane, introducing it as a flat subspace in the bulk, rather than as a dynamical physical object. As the source of the scalar field, we consider a massive point charge moving along a fixed world line localized on the brane. We do not specify the mechanisms of localization of the charge and the ``induced'' kinetic term of the scalar field on the brane, limiting ourselves only to the consideration of the process of radiation of scalar field by the charge.
 
\subsection{Scalar field analog of DGP model}

The action of a scalar DGP model living in a five-dimensional Minkowski space and interacting with a point charge has the form \cite{Dvali:2000hr}
\begin{multline}
\label{eq:sc_DGP_act}
S = M_{5}^{3} \int d^4x dy \, \partial^M \varphi (x;y) \partial_M \varphi (x;y) + M_{4}^{2} \int d^4x dy \, \delta(y) \, \partial^{\mu} \varphi(x;y) \partial_{\mu} \varphi(x;y) - \\ - \int d\tau \, (m + g \varphi(z) ) \sqrt{\eta_{AB} \dot{z}^A \dot{z}^B}, \quad \dot{z}^M = \frac{d z^M}{d\tau},
\end{multline}
where the first term is the analogue of the five-dimensional Einstein-Hilbert action in the bulk, and the second is the analogue of the four-dimensional Einstein-Hilbert term induced on the brane. Here $M_{5}$ and $M_{4}$ are the five-dimensional and four-dimensional Planck masses correspondingly, $m$ and $g$ are the mass and scalar charge of the particle, respectively. We have chosen such a non-canonical normalization of the field, so that in what follows the ratio of two Planck masses will determine the characteristic crossover scale of the theory. The fixed world line of the charge $z^M(\tau)$ is parametrized by its own proper time. The metric of the Minkowski space has the form $\eta_{MN} = {\rm diag} (1,-1,\ldots,-1)$, uppercase Latin indices range over $A=\overline{0,4}$, lowercase Greek the indices enumerate the coordinates on the brane $\mu = \overline{0,3}$, and the coordinate along the extra dimension in the bulk is denoted as $x^4 \equiv y$. The brane is introduced  as a subspace $y=0$ of the complete five-dimensional spacetime, on which an additional kinetic term is localized, and it is not a dynamical object.

The field equation of motion following from the action \eqref{eq:sc_DGP_act} has the form
\begin{equation}
\label{eq:DGP_in_EoM}
M_{5}^{3} \, {^{(5)}}\square \varphi + M_{4}^{2} \, \delta(y) {^{(4)}}\square \varphi = - \frac{g}{2} \int d\tau \, \delta^{(5)}(x-z),
\end{equation}
where ${^{(5)}}\square = \eta^{AB} \partial_A \partial_B$ and ${^{(4)}}\square = \eta^{\alpha\beta} \partial_\alpha \partial_\beta$. Assuming that the charge is localized on the brane $z^M(\tau) = \delta^M_\mu z^\mu(\tau)$, we rewrite the source on the right side of Eq. \eqref{eq:DGP_in_EoM} as
\begin{equation}
\label{eq:source_def}
- \frac{g}{2} \int d\tau \, \delta^{(5)}(x-z) = - \frac{1}{2} j(x) \delta(y), \quad j(x) = g \int d\tau \, \delta^{(4)}(x-z),
\end{equation}
where $j(x)$ is the standard four-dimensional scalar charge current on the brane. Also, introducing the characteristic mass and length scales of the theory as ratios of two Planck masses
\begin{equation}
m_c = \frac{M_{5}^{3}}{M_{4}^{2}}, \quad r_c = \frac{1}{m_c},
\end{equation}
we rewrite the Eq. \eqref{eq:DGP_in_EoM} as
\begin{equation}
\label{eq:DGP_pract_EoM}
{^{(5)}}\square \varphi + r_c \, \delta(y) {^{(4)}}\square \varphi = - \frac{1}{2M_{5}^{3}} j(x) \delta(y).
\end{equation}
In what follows, we consider the Eq. \eqref{eq:DGP_pract_EoM} as the field equation of motion, which determines the radiation from a point charge.

\subsection{Effective four-dimensional field on the brane}

Since in what follows we will be interested in the dynamics of scalar radiation on the brane, we construct the effective four-dimensional equation of motion of the field on the brane, as well as its effective four-dimensional energy-momentum tensor localized on the brane.

Let us start with the effective field equation on the brane. Our derivation differs from those presented in \cite{Hinterbichler:2011tt,deRham:2014zqa,Luty:2003vm} and is intended to clarify physical meaning of the normal branch of the DGP model. We integrate the Eq. \eqref{eq:DGP_pract_EoM} along a coordinate in the bulk over a small interval around the brane $y \in (-\epsilon, \epsilon),\, \epsilon \to +0$ arriving at the equation
\begin{equation}
{^{(4)}}\square \varphi \vert_{y=0} - m_c \lbrack \partial_y \varphi \rbrack_{y=0} = - \frac{1}{2M_{4}^{2}} j(x),
\end{equation}
where $\lbrack \ldots \rbrack_{y=0}$ denotes the jump of the corresponding quantity on the brane. To clarify the four-dimensional meaning of the jump of the field derivative, we Fourier transform the scalar field with respect to the coordinates on the brane
\begin{equation}
\varphi(x;y) = \int \frac{d^4p}{(2\pi)^4} \, e^{-i p_\mu x^\mu} \tilde{\varphi}(p;y).
\end{equation}
Substituting this into the bulk equation of motion
\begin{equation}
{^{(5)}}\square \varphi = 0, \quad y \neq 0
\end{equation}
and solving the resulting equation, we find the general expression for the Fourier transform
\begin{equation}
\label{eq:Four_gen}
\tilde{\varphi}(p;y) =
\begin{cases}
A(p) e^{i \sqrt{p_{\mu}^{2}} y} + B(p) e^{- i \sqrt{p_{\mu}^{2}} y}, \quad y>0 \\
C(p) e^{i \sqrt{p_{\mu}^{2}} y} + D(p) e^{- i \sqrt{p_{\mu}^{2}} y}, \quad y<0.
\end{cases}
\end{equation}
The coefficients $A(p),\ldots,D(p)$ can be chosen in various ways, leading to the choice of the normal or self-accelerating branch of the DGP model. For this we use two physical conditions: we require that the field is continuous on the brane
\begin{equation}
\varphi \vert_{y\to-0} = \varphi \vert_{y\to+0},
\end{equation}
and, also, that the field in the bulk is a superposition of plane waves propagating away from the brane -- radiation boundary condition in the bulk
\begin{equation}
\varphi \sim
\begin{cases}
{\displaystyle \int} e^{-i \omega t + i k y}, \quad y>0 \\
{\displaystyle \int} e^{- i \omega t - i k y}, \quad y<0,
\end{cases}
\end{equation}
which corresponds to the absence of field sources in the bulk. These conditions constrain the coefficients as follows:
\begin{equation}
B(p)=C(p)=0, \quad A(p)=D(p)\equiv\tilde{\varphi}(p).
\end{equation}
For uniqueness of operations with the Fourier transform \eqref{eq:Four_gen} it is necessary to give the definition of the square root in the complex plane (this, also, affects the choice of the DGP model branch). We choose it with a cut along the positive real axis
\begin{align}
\label{eq:SQRT_def_1}
& \sqrt{\rho e^{i\alpha}} = \sqrt{\rho} e^{i\alpha/2}, \\
\label{eq:SQRT_def_2}
& \rho \in \mathbb{R}^{+}, \quad \alpha \in ( 0, 2\pi ).
\end{align}
As a result, by use of the Eq. \eqref{eq:Four_gen} we can formally represent the jump of the field derivative on the brane as
\begin{equation}
\lbrack \partial_y \varphi \rbrack_{y=0} = 2 \sqrt{{^{(4)}}\square} \int \frac{d^4p}{(2\pi)^4} \, e^{-i p_\mu x^\mu} \tilde{\varphi}(p) = 2 \sqrt{{^{(4)}}\square} \left. \varphi \right \vert_{y=0}.
\end{equation}
Then the effective four-dimensional equation of motion of the field on the brane takes the form
\begin{equation}
\label{eq:4D_eff_EoM}
{^{(4)}}\square \varphi \vert_{y=0} - M_c \sqrt{{^{(4)}}\square} \varphi \vert_{y=0} = - \frac{1}{2M_{4}^{2}} j(x),
\end{equation}
where $M_c = 2 m_c$. Given the definition the square root of the d'Alembert operator in terms of the Fourier transform, we obtain the scalar field equation on a brane with a variable mass depending on the momentum.
 
Let us now find the effective four-dimensional energy-momentum tensor of the field localized on the brane. We start with the canonical energy-momentum tensor in the bulk (it is enough to find only its part outside of the world line)
\be
T_{MN} = 2 M_{4}^{2} \, \delta(y) \left \lbrack \eta_{MA} \delta_{\alpha}^{A} \partial^\alpha \varphi \partial_N \varphi - \frac{1}{2} \eta_{MN} \partial^\alpha \varphi \partial_\alpha \varphi \right \rbrack + 2 M_{5}^{3} \left \lbrack \partial_M \varphi \partial_N \varphi - \frac{1}{2} \eta_{MN} \partial^A \varphi \partial_A \varphi \right \rbrack.
\ee
Projecting it onto the brane, we obtain its effective four-dimensional part as
\be
\label{eq:project_EMT}
\delta^{M}_{\mu} \delta^{N}_{\nu} T_{MN} = 2 M_{4}^{2} \, \delta(y) \left \lbrack \partial_\mu \varphi \partial_\nu \varphi - \frac{1}{2} \eta_{\mu\nu} \partial^\alpha \varphi \partial_\alpha \varphi \right \rbrack + 2 M_{5}^{3} \left \lbrack \partial_\mu \varphi \partial_\nu \varphi - \frac{1}{2} \eta_{\mu\nu} \partial^A \varphi \partial_A \varphi \right \rbrack.
\ee
Integrating Eq. \eqref{eq:project_EMT} over a small interval around the brane $y \in (-\epsilon, \epsilon),\, \epsilon \to +0$, we find the effective four-dimensional field energy-momentum tensor localized on the brane
\begin{equation}
\label{eq:4D_eff_EMT}
\Theta_{\mu\nu} \equiv \int_{-\epsilon}^{\epsilon} dy \, \delta^M_\mu \delta^N_\nu T_{MN} = 2 M_{4}^{2} \left. \left \lbrack \partial_\mu \varphi \partial_\nu \varphi - \frac{1}{2} \eta_{\mu\nu} \partial^\alpha \varphi \partial_\alpha \varphi \right \rbrack \right \vert_{y=0}.
\end{equation}
Further, it is the energy-momentum tensor $\Theta_{\mu\nu}$ that will be used to calculate the effective four-dimensional scalar radiation energy flux through the 2-sphere on the brane.

\subsection{Infrared transparency of the bulk -- two-dimensional example}

Before proceeding to the calculation of the retarded field, let's consider a simplified model that allows to better understand the behavior of the field in the bulk and the role of the brane in our construction. Here we follow the setup of   \cite{Brown:2016gwv}.

Consider a two-dimensional scalar DGP model where the only spatial dimension corresponds to a coordinate in the bulk -- the brane is reduced to a single point $y=0$. The equation of motion of the free field in such a model is written as
\begin{equation}
\label{eq:2D_DGP_EoM}
(\partial_{t}^{2} - \partial_{y}^{2}) \varphi + r_c \, \delta(y) \partial_{t}^{2} \varphi = 0.
\end{equation}
Requiring continuity of the field on the brane, we split the Eq.   \eqref{eq:2D_DGP_EoM} into the bulk equation and the matching conditions on the brane, by analogy with the quantum mechanical problem of particle in the $\delta$-function potential,
\begin{align}
\label{eq:2D_bulk_EoM}
& (\partial_{t}^{2} - \partial_{y}^{2}) \varphi = 0, \quad y \neq 0 \\
\label{eq:2D_mat_cond_1}
& \varphi \vert_{y\to-0} = \varphi \vert_{y\to+0}, \\
\label{eq:2D_mat_cond_2}
& - \lbrack \partial_y \varphi \rbrack_{y=0} + r_c \, \partial_{t}^{2} \varphi \vert_{y=0} = 0.
\end{align}
The general solution of Eq. \eqref{eq:2D_bulk_EoM} in the bulk in the form of plane monochromatic waves reads
\begin{equation}
\varphi(t;y) =
\begin{cases}
A(\omega) e^{-i\omega(t-y)} + B(\omega) e^{-i\omega(t+y)}, \quad y<0 \\
D(\omega) e^{-i\omega(t-y)} + C(\omega) e^{-i\omega(t+y)}, \quad y>0.
\end{cases}
\end{equation}
We choose the coefficients in such a way that our solution corresponds to a plane wave incident on the brane from the left, $y<0$, and contains a reflected and a transmitted parts
\begin{equation}
A(\omega) = \frac{1}{\sqrt{2\pi}}, \quad B(\omega) = \frac{R(\omega)}{\sqrt{2\pi}}, \quad C(\omega)=0, \quad D(\omega) = \frac{T(\omega)}{\sqrt{2\pi}}.
\end{equation}
The remaining coefficients are obtained from the matching conditions \eqref{eq:2D_mat_cond_1} and \eqref{eq:2D_mat_cond_2}
\begin{equation}
T(\omega) = \frac{2i}{2i - \omega r_c}, \quad R(\omega) = \frac{\omega r_c}{2i - \omega r_c}.
\end{equation}
As a result, we find the reflection and transmission coefficients for plane monochromatic waves as
\begin{equation}
\label{eq:Trans_coef}
|T(\omega)|^2 = \frac{M_{c}^{2}}{M_{c}^{2} + \omega^2}, \quad |R(\omega)|^2 = \frac{\omega^2}{M_{c}^{2} + \omega^2}.
\end{equation}
Thus, we find that the brane is transparent for waves of frequency $\omega \lesssim M_c$ and opaque for waves of frequency $\omega \gtrsim M_c$. Further, we will see that it results into the fact that only the low frequency signals emitted into the brane are able to leak from it into the extra dimension, while the high frequency ones are quasi-localised on the brane, in accordance with the infrared transparency of the bulk in the DGP model \cite{Dvali:2000xg}.

\subsection{Rohrlich-Teitelboim approach to radiation}

We generalise the Rohrlich-Teitelboim approach to radiation \cite{Rohrlich1961,rohrlich2007,Teitelboim1970} (see, also, \cite{Kosyakov1992,Galtsov:2004uqu,Spirin2009,Galtsov:2010tny}) to extract the emitted part of the DGP scalar field \eqref{eq:sc_DGP_act} and calculate its effective four-dimensional radiation power into the brane. This approach uses certain covariantly defined quantities, whose definitions we recall briefly. Consider the point particle moving along the world line $z^M(\tau)$, parametrized by the proper time $\tau$, with the velocity $v^M=dz^{M}/d\tau$ in the $D$-dimensional Minkowski spacetime. The observation point coordinates are denoted as $x^M$. Assume the observation point to be a top of the past light cone and denote the intersection point of the cone with the particle's world line as $z^M(\hat{\tau}) \equiv \hat{z}^M$. The corresponding moment of proper time $\hat{\tau}$ is called the retarded proper time and is defined by equation
\begin{equation}
\label{eq:ret_prop_time_def}
(x^M - \hat{z}^M)^2 = 0, \quad x^0 \geq \hat{z}^0.
\end{equation}
In what follows, all the hatted quantities correspond to this moment. Based on this, we introduce three spacetime vectors: a null vector $\hat{X}^M = x^M - \hat{z}^M$ directed from the retarded point of the world line to the observation point, a spacelike vector $\hat{u}^M$ orthogonal to the particle's velocity at the retarded proper time, and a null vector $\hat{c}^M = \hat{u}^M + \hat{v}^M$ aligned with $\hat{X}^M$. According to these definitions, we have
\begin{equation}
\hat{X}^2 = 0, \quad (\hat{u}\hat{v}) = 0, \quad \hat{u}^2 = - \hat{v}^2 = -1, \quad \hat{c}^2 = 0,
\end{equation}
where $(\hat{u}\hat{v}) \equiv \hat{u}^A\hat{v}_A$. Using these vectors we introduce the Lorentz-invariant distance $\hat{\rho}$ as the scalar product 
\begin{equation}
\label{eq:Lor-inv_dist_def}
\hat{\rho} \equiv (\hat{v}\hat{X}), \quad \hat{X}^M = \hat{\rho} \hat{c}^M.
\end{equation}
It is equal to the spatial distance in the Lorentz frame comoving with the particle at the retarded proper time. If the particle moves inside the compact region of space, then Lorentz-invariant distance $\hat{\rho}$ is equivalent to the spatial distance $r = |\mathbf{x}|$ when the observation point is far from this region
\begin{equation}
\hat{\rho} \to r, \quad r \gg |\mathbf{z}|.
\end{equation}

In the Rohrlich-Teitelboim approach, it is the Lorentz-invariant distance $\hat{\rho}$ that is used in the long-range expansion of tensors for the definition of the wave zone. Namely, the radiation is determined by the most long-range part of the on-shell energy-momentum tensor expansion in the inverse powers of $\hat{\rho}$. In $D$ dimensions, the retarded field's on-shell energy-momentum tensor expands as \cite{Teitelboim1970,Kosyakov1992,Kosyakov1999,Galtsov:2004uqu,Kosyakov:2008wa,Spirin2009}
\begin{align}
&T^{MN} = T^{MN}_{\rm Coul} + T^{MN}_{\rm mix} + T^{MN}_{\rm rad} \\
&T^{MN}_{\rm Coul} \sim \frac{A^{MN}}{\hat{\rho}^{2D-4}}, \quad T^{MN}_{\rm mix} \sim \frac{B^{MN}}{\hat{\rho}^{2D-5}} + \ldots + \frac{C^{MN}}{\hat{\rho}^{D-1}}, \quad T^{MN}_{\rm rad} \sim \frac{D^{MN}}{\hat{\rho}^{D-2}}.
\end{align}
Here, the first term $T^{MN}_{\rm Coul}$ is the energy-momentum tensor of the deformed Coulomb-like part of the retarded field. The second one is the mixed part, which consists of more than one term for $D>4$ and is absent in $D=3$. The most long-range part $T^{MN}_{\rm rad}$ of the on-shell energy-momentum tensor expansion has the properties allowing to associate it with the radiation energy-momentum:
\begin{itemize}
\item
it is separately conserved $\partial_M T^{MN}_{\rm rad} = 0$, corresponding to its dynamical independence from the other parts;
\item
it is proportional to the direct product of two null vectors $T^{MN}_{\rm rad} \sim \hat{c}^M \hat{c}^N$, corresponding to its propagation exactly with the speed of light $\hat{c}_M T_{\rm rad}^{MN} = 0$;
\item
it falls down as $T^{MN}_{\rm rad} \sim 1/r^{D-2}$ and gives positive definite energy-momentum flux through the distant $(D-2)$-dimensional sphere.
\end{itemize}
Therefore, the radiation power in $D$-dimensions can be computed as the energy flux associated with $T^{MN}_{\rm rad}$ through the distant $(D-2)$-dimensional sphere of radius $r$
\begin{equation}
W_{D} = \int \, T_{\rm rad}^{0i} \, n^{i}\, r^{D-2} \, d\Omega_{D-2}, \quad n^i = x^i/r, \quad i = \overline{1, D-1},
\end{equation}
where $n^i$ is the unit spacelike vector in the direction of observation, and $d\Omega_{D-2}$ is the angular element on the $(D-2)$-dimensional sphere. This structure holds in both even and odd dimensions with the only difference that in odd dimensions the emitted part of the energy-momentum tensor depends on the entire history of the particle's motion preceding the retarded proper time $\hat{\tau}$.

In particular, effective four-dimensional radiation power of the DGP scalar field into the brane in our model is given by the energy flux, associated with the emitted part of the effective four-dimensional energy-momentum tensor on the brane \eqref{eq:4D_eff_EMT}, through the distant 2-sphere on the brane
\begin{equation}
\label{eq:brane_rad_power}
W_{\rm br} = \int \, \Theta^{0i}_{\rm rad} \, n^i \, r^2 \, d\Omega_2, \quad i=\overline{1,3}.
\end{equation}
Note that due to the energy-momentum tensor \eqref{eq:4D_eff_EMT} being the bilinear form of the field derivatives, one can define the emitted part of the retarded field derivative, by analogy with that of the energy-momentum tensor, as its leading $\hat{\rho}$-asymptotic
\begin{equation}
\lbrack \partial_{\mu} \varphi \rbrack^{\rm rad} \sim \frac{1}{\hat{\rho}^{(D-2)/2}}.
\end{equation}

\section{Retarded Green's function of DGP model}\label{III}

The retarded solution of the field equation \eqref{eq:DGP_pract_EoM} is constructed using the retarded Green's function
\begin{equation}
\label{eq:RF_gen}
\varphi(x;y) = - \frac{1}{2M_{5}^{3}} \int d^4x^{\prime} \, G_{\rm ret} (x-x^{\prime};y) \, j(x^{\prime}),
\end{equation}
which satisfies the equations
\begin{align}
\label{eq:5D_GF_eq}
& {^{(5)}}\square G_{\rm ret} (x;y) + r_c \, \delta(y) {^{(4)}}\square G_{\rm ret}(x;y) = \delta^{(4)}(x) \delta(y), \\
& G_{\rm ret}(x;y) = 0, \quad x^0 < 0.
\end{align}
Before calculating the Green's function in five dimensions, let us look at a couple of lower-dimensional examples to clarify the relations between the behavior of the field in the bulk and on the brane and to illustrate the main steps of the calculation.

\subsection{Green's function of 2D DGP model}

In two-dimensional spacetime the equation for the Green's function has the form
\be
\label{eq:1+1_DGP_RGF_eq}
(\partial_{t}^{2} - \partial_{y}^{2}) G(t;y) + r_c \, \delta(y) \partial_{t}^{2} G(t;y) = \delta(t) \delta(y).
\ee
Fourier transforming the Green's function with respect to the temporal coordinate  
\be
\label{eq:2D_GF_FT}
G(t;y) = \int \frac{d\omega}{2\pi} \, e^{- i \omega t} \tilde{G}(\omega;y)
\ee
and substituting it into the Eq. \eqref{eq:1+1_DGP_RGF_eq}, we find equation for the Fourier-image
\be
\partial_{y}^{2} \tilde{G}(\omega;y) + \omega^2 \tilde{G}(\omega;y) + r_c \, \delta(y) \, \omega^2 \tilde{G}(\omega;y) = - \delta(y).
\ee
By analogy with Eqs. (\ref{eq:2D_bulk_EoM}--\ref{eq:2D_mat_cond_2}), we split it into the equation in the bulk and the matching conditions on the brane
\begin{align}
\label{eq:2D_GF_bulk_EoM}
& \partial_{y}^{2} \tilde{G} + \omega^2 \tilde{G} = 0, \quad y \neq 0, \\
\label{eq:2D_GF_mat_cond_1}
& \tilde{G} \vert_{y\to-0} = \tilde{G} \vert_{y\to+0}, \\
\label{eq:2D_GF_mat_cond_2}
& \lbrack \partial_{y} \tilde{G} \rbrack \vert_{y=0} + r_c \, \omega^2 \tilde{G} \vert_{y=0} = -1,
\end{align}
where we required the continuity of the Green's function on the brane. The general solution of Eq. \eqref{eq:2D_GF_bulk_EoM} reads as
\be
\tilde{G}(\omega;y) =
\begin{cases}
A(\omega) e^{- i \omega y} + B(\omega) e^{i \omega y}, \quad y<0, \\
C(\omega) e^{- i \omega y} + D(\omega) e^{i \omega y}, \quad y>0.
\end{cases}
\ee
Imposing on the Fourier-image the radiation boundary condition in the bulk with account for Eq. \eqref{eq:2D_GF_mat_cond_1} we get
\be
B(\omega)=C(\omega)=0, \quad A(\omega)=D(\omega).
\ee
The remaining coefficient is obtained from the Eq. \eqref{eq:2D_GF_mat_cond_2}. As a result, the Fourier-image of the Green's function takes the form 
\be
\tilde{G}(\omega;y) = - \frac{e^{i \omega |y|}}{\omega^2 r_c + 2 i \omega},
\ee
so finally we get 
\be
\label{eq:1+1_DGP_RGF_FT}
G(t;y) = - \int \frac{d\omega}{2\pi} \, \frac{e^{- i \omega (t - |y|)}}{\omega^2 r_c + 2i\omega}.
\ee
The integrand in Eq. \eqref{eq:1+1_DGP_RGF_FT} has two singular points
\be
\omega = 0, \quad \omega = - i M_c.
\ee
To satisfy the retardation condition, we close the integration contour in the upper complex half-plane for $t<|y|$ and in the lower half-plane for $t>|y|$. As a result, the Green's function takes the form
\be
\label{eq:1+1_DGP_RGF}
G(t;y) = \frac{1}{2} \theta(t-|y|) \left \lbrack 1- e^{-M_c (t-|y|)} \right \rbrack.
\ee

\begin{figure}[t]
\center{\includegraphics[width=0.6\linewidth]{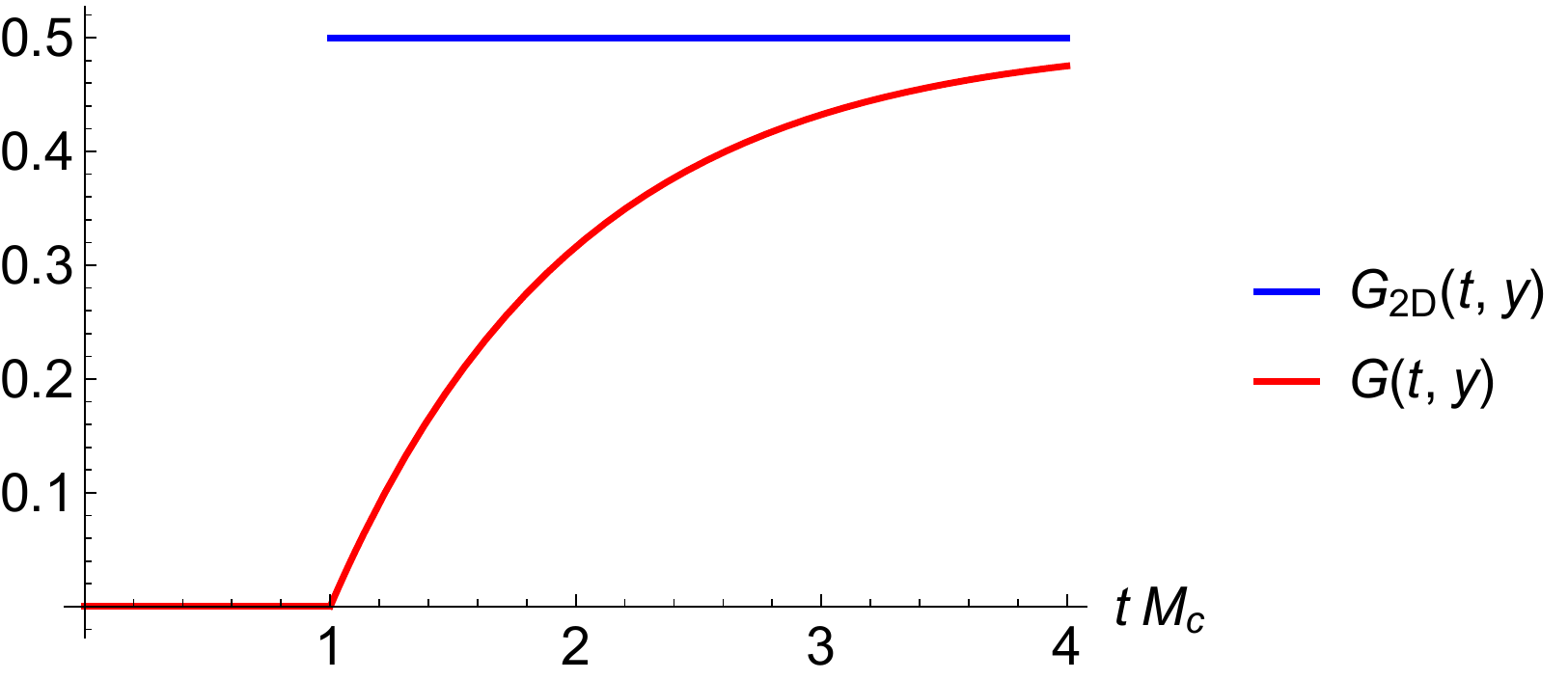}}
\caption{The retarded Green's function of a two-dimensional DGP model and the retarded Green's function of a two-dimensional massless scalar field (here $|y|=M_c^{-1}$).}
\label{fig:1}
\end{figure}

Compare the Eq. \eqref{eq:1+1_DGP_RGF} with the Green's function of a two-dimensional massless scalar field \cite{Ivanenko_book}
\be
\label{eq:1+1_Mink_RGF}
G_{\rm 2D}(t,y) = \frac{1}{2} \theta (t-|y|),
\ee
where we denote the spatial coordinate, also, as $y$. From Fig. \eqref{fig:1} we find that in the two-dimensional scalar DGP model, field perturbations are temporarily quasi-localized on the brane during the time interval $\Delta t \sim M_{c}^{-1}$, in accordance with the metastable nature of the effective graviton on a brane in the DGP model \cite{Gabadadze:2003ck,Gabadadze:2004dq}.

\subsection{Green's function of 3D DGP model}

Here our calculations are similar to those presented in the Ref. \cite{Hinterbichler:2009kq}. In the three-dimensional DGP model, the equation for the Green's function is written as
\be
\label{eq:3D_GF_def}
{^{(3)}}\square G(x;y) + r_c \, \delta(y) {^{(2)}}\square G(x;y) = \delta^{(2)}(x) \delta(y).
\ee
By analogy with the two-dimensional model, Fourier transforming the Green's function with respect to the coordinates on the brane
\be
\label{eq:3D_GF_FT}
G(x;y) = \int \frac{d^2p}{(2\pi)^2} \, e^{-i p_\mu x^\mu} \tilde{G}(p;y), \quad \mu, \nu = \overline{0,1}
\ee
we obtain the equation for the Fourier-image of the Green's function in the bulk and the matching conditions on the brane
\begin{align}
\label{eq:3D_GF_bulk_EoM}
& \partial_{y}^{2} \tilde{G} + p_{\mu}^{2} \tilde{G} = 0, \quad y \neq 0 \\
\label{eq:3D_GF_mat_cond_1}
& \tilde{G} \vert_{y\to-0} = \tilde{G} \vert_{y\to+0}, \\
\label{eq:3D_GF_mat_cond_2}
& \lbrack \partial_y \tilde{G} \rbrack \vert_{y=0} + r_c \, p_{\mu}^{2} \tilde{G} \vert_{y=0} = -1.
\end{align}
Solving the equation of motion in the bulk \eqref{eq:3D_GF_bulk_EoM} and using the matching conditions on the brane \eqref{eq:3D_GF_mat_cond_1} and \eqref{eq:3D_GF_mat_cond_2} together with the radiation boundary condition in the bulk, we arrive at the following Fourier integral for the Green's function
\be
\label{eq:2+1_RGF_FI}
G(x;y) = - \int \frac{d^2p}{(2\pi)^2} \, \frac{e^{-i p_\mu x^\mu} e^{i \sqrt{p_{\mu}^{2}} |y|}}{r_c \, p_{\mu}^{2} + 2i \sqrt{p_{\mu}^{2}}},
\ee
where the square root is defined by the equations (\ref{eq:SQRT_def_1}--\ref{eq:SQRT_def_2}). In higher dimensions, the Fourier integral for the Green's function will have a similar form. Further, we will only be interested in the propagation of the field along the brane $y=0$. In this case, the Fourier integral for the Green's function is simplified
\be
G(x;0) = - \frac{M_c}{2} \int \frac{d^2p}{(2\pi)^2} \, \frac{e^{-i \omega t} e^{i p x}}{\omega^2 - p^2 + i M_c \sqrt{\omega^2 - p^2}}, \quad \omega = p^0, \quad p = p^1.
\ee

\begin{figure}[t]
\begin{minipage}[h]{0.49\linewidth}
\center{\includegraphics[width=0.8\linewidth]{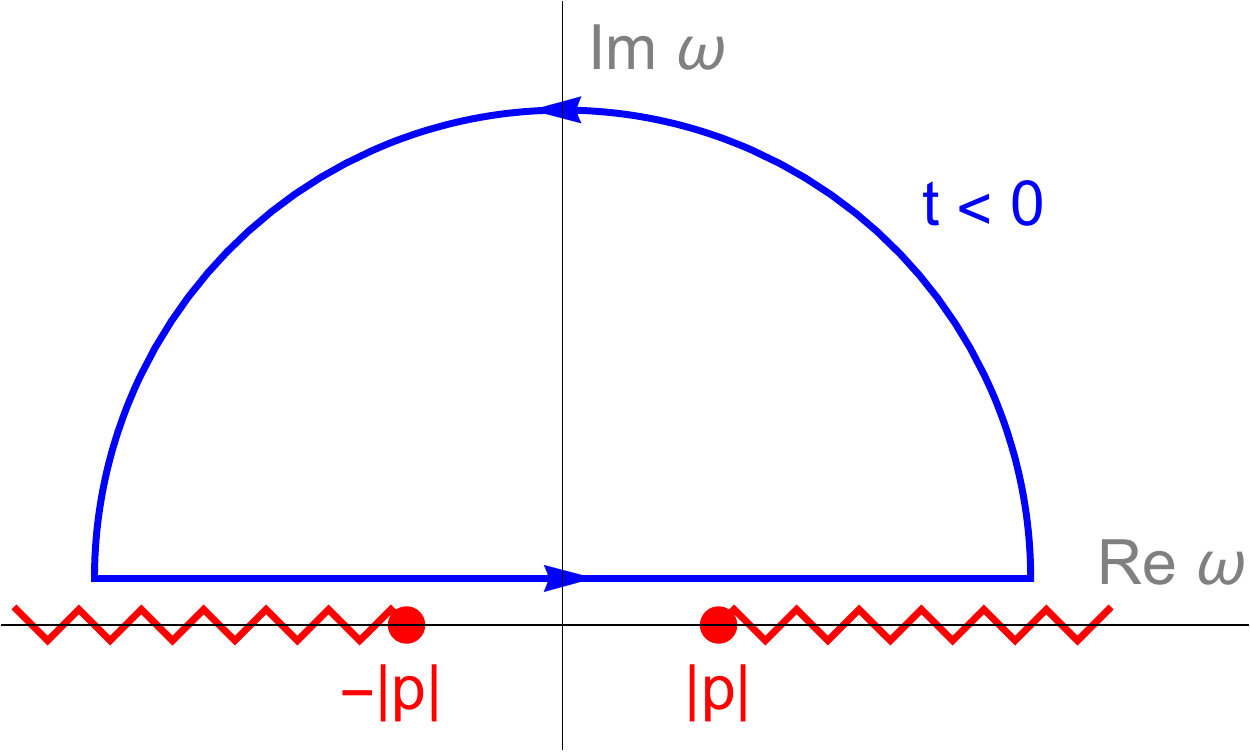}}
\end{minipage}
\hfill
\begin{minipage}[h]{0.49\linewidth}
\center{\includegraphics[width=0.8\linewidth]{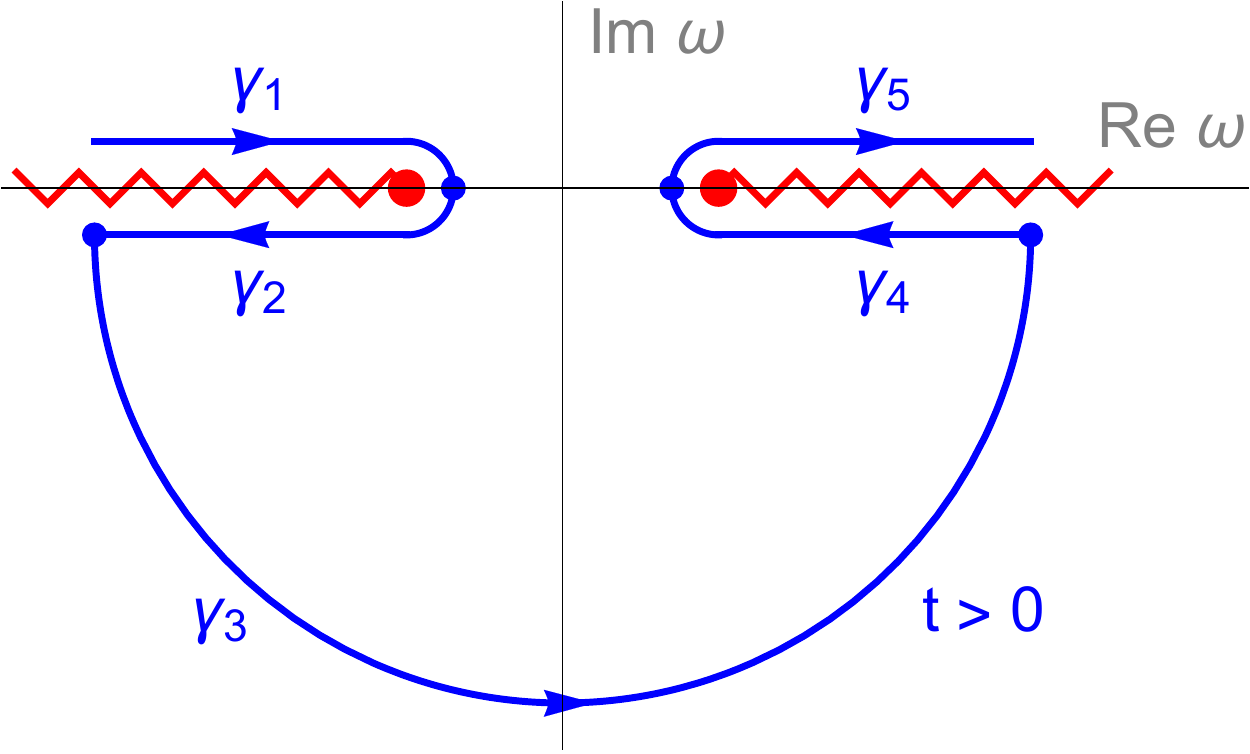}}
\end{minipage}
\caption{Integration contours on the complex $\omega$-plane corresponding to the retardation condition.}
\label{fig:2}
\end{figure}

By virtue of the definition of a square root, there are cuts on the complex $\omega$-plane. In accordance with the Eqs. (\ref{eq:SQRT_def_1}--\ref{eq:SQRT_def_2}), they lie on the real axis, see Fig. \eqref{fig:2},
\be
\omega^2 - p^2 \geq 0 \quad \Longrightarrow \quad \omega \in (-\infty, -|p| \, \rbrack \cup \lbrack \, |p|, \infty).
\ee
When finding the singular points of the integrand, we take into account that, in accordance with the definition of the square root, its imaginary part must be non-negative
\be
{\rm Im} \, \sqrt{z} \geq 0.
\ee
Then, from zeros of the denominator of the integrand
\be
\omega^2 - p^2 + i M_c \sqrt{\omega^2 - p^2} = 0 \quad \Longrightarrow \quad \omega = \pm |p|, \quad \sqrt{\omega^2 - p^2} = - i M_c,
\ee
we find that the first two singular points are the beginnings of the cuts, and the other two do not lie on the chosen Riemann sheet, in accordance with the normal branch of the DGP model \cite{Gabadadze:2004dq}. Thus the integrand has no singular points except the cuts in the complex $\omega$-plane.

Since the cuts lie on the real axis, we have to shift the integration contour for $\omega \in (-\infty,\infty)$ integral infinitesimally above or below the cuts. The direction of the shift is chosen in such a way that when the contour is closed, the resulting Green's function satisfies the retardation condition. This condition is satisfied by the upward shift of the contour over the cuts, see Fig. \eqref{fig:2}. Indeed, shifting the contour upward and closing it in the upper half-plane for $t<0$, we obtain
\be
G(x;0) \sim - \int {\rm Res} \, \frac{e^{-i\omega t}}{\omega^2 - p^2 + i M_c \sqrt{\omega^2 - p^2}} = 0, \quad t<0.
\ee
However, the cuts do not allow us to close the contour in the lower half-plane for $t>0$. Therefore, we deform it so that it encloses the cuts on both sides, see Fig. \eqref{fig:2}. The contribution of the arc resulting from such a deformation of the contour and lying in the lower half-plane vanishes at $t>0$. As a result, the retarded Green's function is given by the integral
\be
G(x;0) = - \frac{M_c}{2} \, \theta(t) \left( \int_{\gamma_1} - \int_{\gamma_{2}^{-}} - \int_{\gamma_{4}^{-}} + \int_{\gamma_{5}} \right) \frac{d^2p}{(2\pi)^2} \, \frac{e^{-i\omega t} e^{ipx}}{\omega^2 - p^2 + i M_c \sqrt{\omega^2 - p^2}},
\ee
where index $\gamma_{n}^{-}$ denotes the integration over the contour $\gamma_n$ in the opposite direction. On the upper and lower sides of the cuts, the square root takes different values
\begin{align}
\label{eq:Compl_cut_1}
& \gamma_1: \quad \sqrt{\omega^2 - p^2} = \xi, \quad \xi^2 \equiv \omega^2 - p^2, \quad \xi \in \mathbb{R}^{-}, \\
& \gamma_{2}^{-}: \quad \sqrt{\omega^2 - p^2} = - \xi, \quad \xi^2 \equiv \omega^2 - p^2, \quad \xi \in \mathbb{R}^{-}, \\
& \gamma_{4}^{-}: \quad \sqrt{\omega^2 - p^2} = - \xi, \quad \xi^2 \equiv \omega^2 - p^2, \quad \xi \in \mathbb{R}^{+}, \\
\label{eq:Compl_cut_2}
& \gamma_{5}: \quad \sqrt{\omega^2 - p^2} = \xi, \quad \xi^2 \equiv \omega^2 - p^2, \quad \xi \in \mathbb{R}^{+}.
\end{align}
Here the variable $\xi$ is introduced in such a way that on two cuts together it takes all real values $\xi \in (-\infty,\infty)$. Taking into account the Eqs. (\ref{eq:Compl_cut_1}--\ref{eq:Compl_cut_2}) Green's function is written as
\be
G(x;0) = - \frac{M_c}{8\pi^2} \, \theta(t) \int dp \, e^{ipx} \left( \int_{-\infty}^{-|p|} + \int_{|p|}^{\infty} \right) d\omega \, e^{-i \omega t} \, \Delta \! \left \lbrack \frac{1}{\xi^2 + i M_c \sqrt{\omega^2 - p^2}} \right \rbrack,
\ee
where $\Delta \lbrack \ldots \rbrack$ denotes the jump of the integrand on the corresponding cut -- the value of the function above the cut minus the value below.

\begin{figure}[t]
\center{\includegraphics[width=0.65\linewidth]{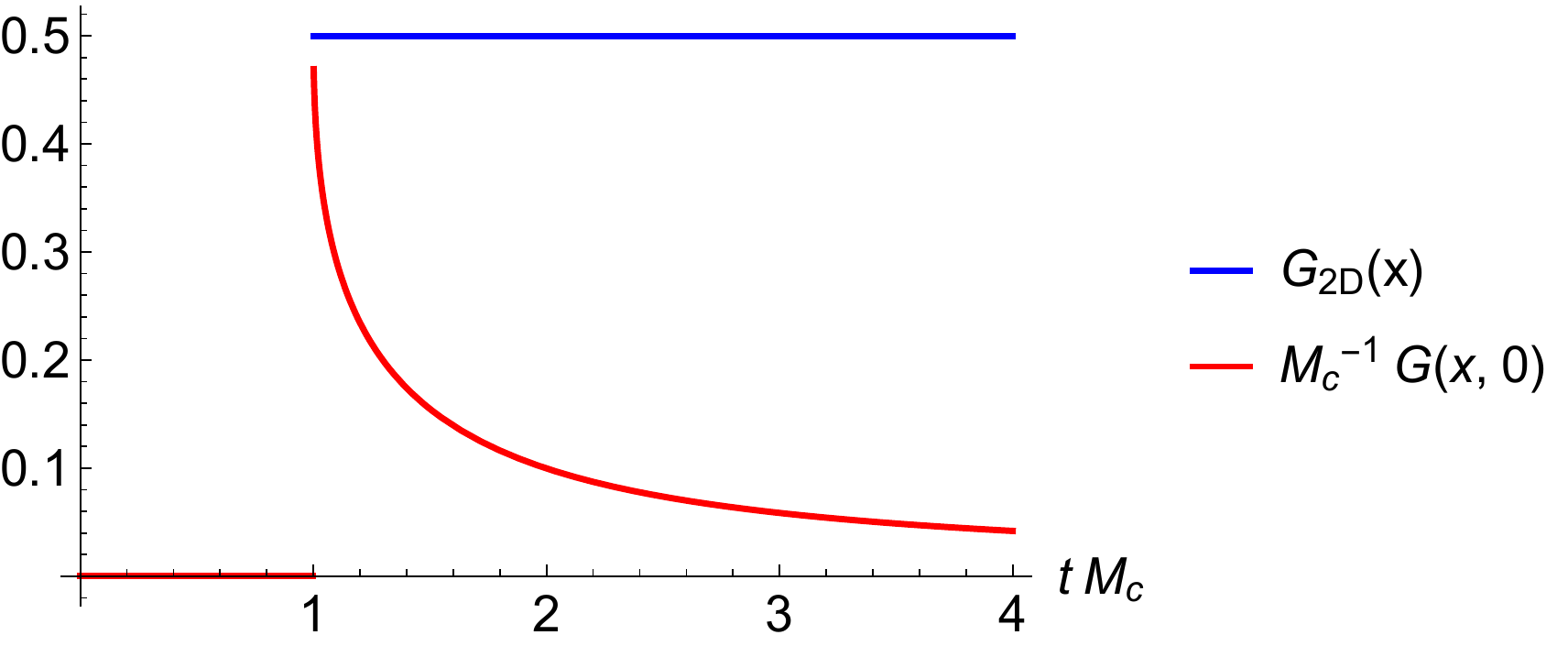}}
\caption{The retarded Green's function on the brane of the three-dimensional DGP model and the retarded Green's function of the two-dimensional massless scalar field (here $|x|=M_c^{-1}$).}
\label{fig:3}
\end{figure}

Calculating the jumps on the cuts by use of the Eqs. (\ref{eq:Compl_cut_1}-\ref{eq:Compl_cut_2}), we arrive at the following integral for the Green's function
\be
\label{eq:3D_GF_xi_int}
G(x;0) = \frac{M_{c}^{2}}{4\pi^2} \, \theta(t) \int dp \, e^{ipx} \int d\xi \, \frac{\sin \sqrt{\xi^2 + p^2}t}{\sqrt{\xi^2 + p^2}} \frac{1}{\xi^2 + M_{c}^{2}},
\ee
where we have taken into account the relation $\omega = {\rm sgn}(\xi) \sqrt{\xi^2 + p^2}$. Due to the sine being even function, the integrand here is single-valued and analytic in the whole complex $\xi$- and $p$-planes. The remaining integrals are easily calculated \cite{zwillinger2014table}, and the Green's function on the brane takes the form
\be
\label{eq:2+1_DGP_RGF_br_exact}
G(x;0) = \frac{M_c}{4} \, \theta(t-|x|) \left \lbrack I_0 ( M_c \sqrt{t^2 - x^2} ) - \mathbb{L}_0 ( M_c \sqrt{t^2 - x^2} ) \right \rbrack,
\ee
where $I_0(x)$ is a modified Bessel function of the first kind and $\mathbb{L}_0(x)$ is a Struve function. Comparing the Eq. \eqref{eq:2+1_DGP_RGF_br_exact} with the Green's function of the two-dimensional massless field \eqref{eq:1+1_Mink_RGF}, we find that the field perturbation on the brane, having reached the observation point at the speed of light, decays with time, in contrast to the case of a two-dimensional massless field, see Fig. \eqref{fig:3}. This decay corresponds to the gradual leakage of the field from the brane, in accordance with the metastable nature of the effective graviton on the brane in the DGP model \cite{Gabadadze:2003ck,Gabadadze:2004dq}. In agreement with the two-dimensional DGP model \eqref{eq:1+1_DGP_RGF}, the field perturbation is quasi-localized on the brane for the time interval $\Delta t \sim M_{c}^{-1}$.

However, Eq. \eqref{eq:2+1_DGP_RGF_br_exact} for the Green's function on the brane appears to be inconvenient for the further computations. Therefore, we rewrite it in another form. We note that in Eq. \eqref{eq:3D_GF_xi_int} the integral over momentum is proportional to the Green's function of a two-dimensional massive field \cite{Ivanenko_book} and $\xi$ plays here the role of effective mass. Thus, redenoting $\xi=\mu$, we rewrite the Green's function on the brane as
\begin{align}
& G(x;0) = \frac{1}{\pi} \int_{0}^{\infty} d\mu \, \rho(\mu) \, G_{\rm 2D}(x|\mu), \quad \rho(\mu) = \frac{M_{c}^{2}}{M_{c}^{2} + \mu^2}, \\
& G_{\rm 2D}(x|\mu) = \frac{1}{2} \, \theta(t) \theta(t^2 - x^2) J_0( \mu \sqrt{t^2 - x^2} ),
\end{align}
where $G_{\rm 2D}(x|\mu)$ is the retarded Green's function of a two-dimensional massive field \cite{Ivanenko_book}. Thus, the field on the brane is a set of two-dimensional massive Kaluza-Klein modes with masses running over a continuous range of values, by analogy with the effective graviton on the brane in the DGP model \cite{Dvali:2001gm,Dvali:2001gx,Gabadadze:2003ck,Gabadadze:2004dq}. The spectral function $\rho(\mu)$ in the integral over the Kaluza-Klein masses determines the characteristic range of modes that contribute to the field perturbations on the brane.

\subsection{Green's function of 5D DGP model}

Let us now find the retarded Green's function of the five-dimensional scalar DGP model. By analogy with the Eqs. (\ref{eq:3D_GF_FT}--\ref{eq:3D_GF_mat_cond_2}), we Fourier transform the Green's function over the coordinates on the brane, substitute it into the Eq. \eqref{eq:5D_GF_eq} and split the resulting equation into the equation for the Fourier-image in the bulk and the matching conditions on the brane. From the resulting system of equations, we find the Fourier integral for the Green's function similar to the Eq. \eqref{eq:2+1_RGF_FI}
\be
G_{\rm ret}(x;y) = - \int \frac{d^4p}{(2\pi)^4} \, \frac{e^{-i p_\mu x^\mu} e^{i \sqrt{p_{\mu}^{2}} |y|}}{r_c \, p_{\mu}^{2} + 2i \sqrt{p_{\mu}^{2}}}.
\ee
We are only interested in the Green's function on the brane. In this case, the Fourier integral reduces to
\be
\label{eq:4+1_RGF_FI_br}
G_{\rm ret}(x;0) = - \int \frac{d^4p}{(2\pi)^4} \, \frac{e^{-i \omega t} e^{i \mathbf{p} \mathbf{x} }}{r_c (\omega^2 - \mathbf{p}^2) + 2i \sqrt{\omega^2 - \mathbf{p}^2}}.
\ee
One can show that the Eq. \eqref{eq:4+1_RGF_FI_br} coincides with the Fourier integral for the Green's function of the effective four-dimensional equation of motion of the field on the brane \eqref{eq:4D_eff_EoM}.

The integral \eqref{eq:4+1_RGF_FI_br} is calculated by analogy with the three-dimensional Green's function considered in the previous section. The only difference here is the triple integral over the spatial components of momentum in Eq. \eqref{eq:3D_GF_xi_int}, which is easily calculated \cite{Ivanenko_book,zwillinger2014table}. As a result, we find the following Kaluza-Klein decomposition for the retarded Green's function on the brane
\begin{align}
\label{eq:4+1_RGF_KKI_1}
& G_{\rm ret}(x;0) = \frac{1}{\pi} \int_{0}^{\infty} d\mu \, \rho(\mu) \, G_{\rm 4D}(x|\mu), \quad \rho(\mu) = \frac{M_c^2}{\mu^2 + M_c^2} \\
\label{eq:4+1_RGF_KKI_2}
& G_{\rm 4D}(x|\mu) = \frac{\theta(t)}{2\pi} \left \lbrack \delta(x^2) J_0(\mu \sqrt{x^2}) - \frac{1}{2} \frac{\theta(x^2)}{\sqrt{x^2}} \mu J_1(\mu \sqrt{x^2}) \right \rbrack,
\end{align}
where $G_{\rm 4D}(x|\mu)$ is the retarded Green's function of the four-dimensional massive scalar field \cite{Ivanenko_book}. Calculating the integral over Kaluza-Klein masses \cite{zwillinger2014table}, we find the explicit form of the Green's function on the brane
\begin{multline}
\label{eq:4+1_RGF_expl}
G_{\rm ret}(x;0) = \frac{M_c}{4\pi} \, \theta(t) \bigg \lbrace \delta(x^2) \left \lbrack I_0(M_c \sqrt{x^2}) - \mathbb{L}_0(M_c \sqrt{x^2}) \right \rbrack + \\ + \frac{M_c}{2} \frac{\theta(x^2)}{\sqrt{x^2}} \left \lbrack I_1(M_c \sqrt{x^2}) - \frac{1}{2} \mathbb{L}_{-1}(M_c \sqrt{x^2}) - \frac{1}{2} \mathbb{L}_1(M_c \sqrt{x^2}) -  \frac{1}{\pi} \right \rbrack \bigg \rbrace.
\end{multline}
However, the Kaluza-Klein decomposition of the Green's function \eqref{eq:4+1_RGF_KKI_1} is more convenient in the further computations. Note that, in accordance with the Huygens principle violation in the five-dimensional bulk of the DGP model, the resulting Green's function \eqref{eq:4+1_RGF_expl} is localized not only on the light cone, but also inside it. Correctness of the Eqs. (\ref{eq:4+1_RGF_KKI_1}--\ref{eq:4+1_RGF_KKI_2}) is verified by the calculation of the field of static charge (see Appendix A). The obtained field coincides with the expression found in the normal branch of the DGP model \cite{Dvali:2000hr}, which confirms its connection with the radiation boundary condition in the bulk.

As a result, the retarded field on the brane, in accordance with Eq. \eqref{eq:RF_gen} reads as
\be
\label{eq:ret_field_br}
\varphi(x;0) = - \frac{1}{2 \pi M_{5}^{3}} \int_{0}^{\infty} d\mu \, \rho(\mu) \int d^4x' \, G_{\rm 4D}(x-x'|\mu) \, j(x').
\ee
For convenience of further calculations, we rewrite it as a continuous spectrum of four-dimensional massive Kaluza-Klein scalar fields with the sources $j(x)$
\begin{align}
\label{eq:RF_mass_decomp_1}
& \varphi(x;0) = - \frac{1}{2\pi M_{5}^{3}} \int_{0}^{\infty} d\mu \, \rho(\mu) \, \psi(x|\mu), \\
\label{eq:RF_mass_decomp_2}
& {^{(4)}}\square \psi(x|\mu) + \mu^2 \psi(x|\mu) = j(x), \\
\label{eq:RF_mass_decomp_3}
& \psi(x|\mu) = \int d^4x' \, G_{\rm 4D}(x-x'|\mu) \, j(x').
\end{align}

To calculate the effective four-dimensional energy flux of scalar radiation through a 2-sphere on the brane, it is enough to find the radiated part of derivative of the field on the brane, due to its effective four-dimensional energy-momentum tensor on the brane \eqref{eq:4D_eff_EMT} being just the bilinear form of field derivatives. In turn, the radiated part of derivative of the field on the brane is determined by the radiated part of derivative of the four-dimensional massive scalar field
\be
\label{eq:4+1_DGP_rad_gen}
\lbrack \partial_\mu \varphi(x;0) \rbrack^{\rm rad} = - \frac{1}{2\pi M_{5}^{3}} \int_{0}^{\infty} d\mu \, \rho(\mu) \, \lbrack \partial_\mu \psi(x|\mu) \rbrack^{\rm rad}.
\ee
As a result, the problem of radiation in the scalar DGP model is reduced to the problem of radiation of a four-dimensional massive scalar field. In turn, the radiated part of derivative of the four-dimensional massive field can be found due to its connections with massless fields in dimensions four and five, whose emitted parts are well-known \cite{Spirin2009,Galtsov:2020hhn}.

\section{Radiated part of 4D massive field}\label{IV}

The complete solution of the problem of radiation of a four-dimensional massive field (\ref{eq:RF_mass_decomp_2}--\ref{eq:RF_mass_decomp_3}) seems to be a non-trivial task of generalization of the Rohrlich-Teitelboim approach, since here even a free field does not propagate exactly at the speed of light. Actually, a new geometric definition of the emitted four-momentum would be required. However, the concept of the radiated part of the {\em field derivative} can be directly extrapolated to the massive theory due to its connections with massless fields in dimensions four and five.

\subsection{Dimensional reduction of 5D massless field}

Consider a five-dimensional massless scalar field with the source $j(x)$ on a flat 3-brane. Its equation of motion has the form
\be
\label{eq:5D_WE}
{^{(5)}}\square \Psi(x;y) = j(x) \delta(y).
\ee
We Fourier transform the field over the coordinate in the extra dimension
\be
\Psi(x;y) = \int \frac{d\mu}{2\pi} \, e^{i \mu y} \, \tilde{\Psi}(x;\mu),
\ee
where we denote the component of five-momentum in the bulk as $\mu$. Substituting this into Eq. \eqref{eq:5D_WE}, we obtain the effective equation of motion for the Fourier-image
\be
\label{eq:4D_eff_MWE}
{^{(4)}}\square \tilde{\Psi}(x;\mu) + \mu^2 \tilde{\Psi}(x;\mu) = j(x),
\ee
coinciding with the equation for a four-dimensional massive field \eqref{eq:RF_mass_decomp_2}. Retarded solution of the Eq. \eqref{eq:5D_WE} can be constructed from the retarded solution of the effective equation for the Fourier-image \eqref{eq:4D_eff_MWE}, which has the form
\be
\tilde{\Psi}(x;\mu) = \int d^4x' \, G_{\rm 4D}(x-x'|\mu) \, j(x') = \psi(x|\mu),
\ee
where $\psi(x|\mu)$ is defined by the Eqs. (\ref{eq:RF_mass_decomp_2}--\ref{eq:RF_mass_decomp_3}). Due to the four-dimensional massive Green's function \eqref{eq:4+1_RGF_KKI_2} being even function of the field mass, the retarded solution of the Eq. \eqref{eq:5D_WE} is written as
\be
\Psi(x;y) = \frac{1}{\pi} \int_{0}^{\infty} d\mu \, \cos{ \mu y} \, \psi(x|\mu).
\ee
In particular, the field on the brane takes the form
\be
\label{eq:dim_red}
\Psi(x;0) = \frac{1}{\pi} \int_{0}^{\infty} d\mu \, \psi(x|\mu),
\ee
which is analogous to the field on the brane in the DGP model \eqref{eq:RF_mass_decomp_1}, but here the spectral function has simple form $\rho(\mu)=1$. Thus, five-dimensional massless field on the brane is represented as a continuous spectrum of four-dimensional massive fields.

Therefore, emitted part of five-dimensional massless field on the brane should, also, have the Kaluza-Klein decomposition into the continuous spectrum of emitted parts of the four-dimensional massive fields
\be
\label{eq:CC_dim_red}
\lbrack \partial_\mu \Psi(x;0) \rbrack^{\rm rad} = \frac{1}{\pi} \int_{0}^{\infty} d\mu \, \lbrack \partial_\mu \psi(x|\mu) \rbrack^{\rm rad}.
\ee
Since the emitted part of the five-dimensional massless field, in accordance with the Rohrlich-Teitelboim approach, is extracted as the leading $\hat{\rho}$-asymptotic of its derivative \cite{Spirin2009,Galtsov:2020hhn}, we conjecture that due to the Eq. \eqref{eq:CC_dim_red} the emitted part of derivative of the four-dimensional massive field can, also, be extracted in a similar way. Below we will demonstrate that the radiated part of the four-dimensional massive field extracted in accordance with the Rohrlich-Teitelboim approach together with Eq. \eqref{eq:CC_dim_red} leads to the correct expression for the radiated part of the five-dimensional massless field.

Note that the Eq. \eqref{eq:CC_dim_red} holds, indeed, in the case of a charge moving inside the brane, since, by virtue of the definitions of retarded covariant quantities (\ref{eq:ret_prop_time_def} -- \ref{eq:Lor-inv_dist_def}), we have
\begin{align}
\label{eq:RCQ_dim_rel_1}
& \hat{\rho}_{\rm 5D} \vert_{y=0} = \hat{v}_M ( x^M - \hat{z}^M ) \vert_{y=0} = \hat{v}_\mu \left( x^\mu - \hat{z}^\mu \right) = \hat{\rho}_{\rm 4D} \equiv \hat{\rho}, \\
\label{eq:RCQ_dim_rel_2}
& \hat{c}_{\rm 5D}^{M} \vert_{y=0} = \frac{1}{\hat{\rho}} (x^M - \hat{z}^M) \vert_{y=0} = \frac{1}{\hat{\rho}} \delta^M_\mu (x^\mu - \hat{z}^{\mu}) = \left \lbrack \hat{c}_{\rm 4D}^{\mu}; 0 \right \rbrack,
\end{align}
where $\hat{\rho}_{\rm 5D}$ and $\hat{c}_{\rm 5D}^{M}$ are covariant retarded quantities defined in the five-dimensional bulk, and $\hat{\ rho}_{\rm 4D}$ and $\hat{c}_{\rm 4D}^{\mu}$ are defined on the brane.

\subsection{Massless limit of 4D field}

Correctness of application of the Rohrlich-Teitelboim approach for the extraction of emitted part of the four-dimensional massive field derivative can be additionally verified by calculation of its massless limit. Indeed, in the massless limit $\mu \to 0$ radiated part of the four-dimensional massive field must coincide with the emitted part of the massless four-dimensional field
\begin{align}
\label{eq:massless_limit_cond}
& \lbrack \partial_\mu \psi(x|\mu) \rbrack^{\rm rad} \xrightarrow{\mu \to 0} \lbrack \partial_\mu \psi(x) \rbrack^{\rm rad}, \\
& {^{(4)}}\square \psi(x|\mu) + \mu^2 \psi(x|\mu) = j(x) \xrightarrow{\mu \to 0} {^{(4)}}\square \psi(x) = j(x).
\end{align}

Let us find the radiated part of the four-dimensional massless field for the field source being one point charge. The retarded field is written as
\begin{align}
\label{eq:4D_ml_ret}
& \psi(x) = \int d^4x' \, G_{\rm 4D}(x-x') \, j(x'), \\
& G_{\rm 4D}(x) = \frac{\theta(t) \delta(x^2)}{2\pi},
\end{align}
where $G_{\rm 4D}(x)$ is the retarded Green's function of a four-dimensional massless field \cite{Ivanenko_book}. Substituting here the source \eqref{eq:source_def}, we arrive at the retarded field in the form
\be
\psi(x) = g \int d\tau \, \frac{\theta(X^0) \delta(X^2)}{2\pi}, \quad X^\mu = x^\mu - z^\mu(\tau).
\ee
Using relation for the $\delta$-function of a complex argument
\be
\label{eq:d-func_comp_arg}
\theta(X^0) \delta(X^2) = \frac{\delta(\tau - \hat{\tau})}{2\hat{\rho}}
\ee
we get a simple expression for the retarded field
\be
\label{eq:4D_RF_massless}
\psi(x) = \frac{g}{4\pi\hat{\rho}}.
\ee

Using the rule of differentiation of the covariant retarded quantities \cite{Kosyakov1992,Kosyakov1999}
\be
\partial_\mu \hat{\rho} = \hat{v}_\mu + \hat{c}_\mu (\hat{\rho} (\hat{a}\hat{c}) - 1), \quad a^\mu = \frac{d^2 z^\mu}{d\tau^2}
\ee
we find derivative of the retarded four-dimensional massless field as
\be
\partial_\mu \psi(x) = - \frac{g(\hat{v}_\mu - \hat{c}_\mu)}{4\pi\hat{\rho}^2} - \frac{g\hat{c}_\mu (\hat{a}\hat{c})}{4\pi\hat{\rho}}.
\ee
In accordance with the Rohrlich-Teitelboim approach, we extract emitted part of the four-dimensional massless field as the leading $\hat{\rho}$-asymptotic of its derivative arriving at
\be
\label{eq:4D_massless_rad}
\lbrack \partial_\mu \psi(x) \rbrack^{\rm rad} = - \frac{g \hat{c}_\mu (\hat{a}\hat{c})}{4\pi \hat{\rho}}.
\ee

\subsection{Emitted part of 5D massless field}

Let us find the emitted part of five-dimensional massless scalar field of a point charge. We will extract the radiated part of four-dimensional massive field by analogy with this calculation.

In five dimensions, the retarded massless field of a point charge reads
\begin{align}
& \Psi(x;y) = \int d^5x' \, G_{\rm 5D}(x-x') j(x'), \quad j(x) = g \int d\tau \, \delta^{(5)}(x-z), \\
& G_{\rm 5D}(x) = \frac{\theta(t)}{2\pi^2} \left \lbrack \frac{\delta(x^2)}{\sqrt{x^2}} - \frac{1}{2} \frac{\theta(x^2)}{(x^2)^{3/2}} \right \rbrack,
\end{align}
where $G_{\rm 5D}(x)$ is the retarded Green's function of the five-dimensional massless field. For convenience, we do not limit the region of charge motion to the brane. By use of the Eqs. (\ref{eq:RCQ_dim_rel_1}--\ref{eq:RCQ_dim_rel_2}) this can be easily done in the final result. Substituting the scalar current into the integral and using the Eq. \eqref{eq:d-func_comp_arg}, we get
\be
\Psi(x;y) = - \frac{g}{4\pi^2} \int_{-\infty}^{\hat{\tau}} d\tau \left \lbrack \frac{1}{(X^2)^{3/2}} - \frac{\delta(\tau - \hat{\tau})}{\hat{\rho} \sqrt{X^2}} \right \rbrack, \quad X^M = x^M - z^M(\tau).
\ee
Note that, by virtue of Eq. \eqref{eq:ret_prop_time_def} each of the terms of the integrand diverges at the upper integration limit. However, all the physical information about the field is contained in the first term, while the local term with the $\delta$-function just eliminates the divergence contained in the first one \cite{Khlopunov:2022ubp,Galtsov:2020hhn,Galtsov:2021zpb}. Introducing a regularizing parameter into the upper integration limit $\hat{\tau} \to \hat{\tau}-\epsilon, \, \epsilon \to +0$ and integrating the first term by parts by use of the relation
\be
\label{eq:5D_int_parts_rel}
\frac{1}{(X^2)^{3/2}} = \frac{1}{(vX)} \frac{d}{d\tau} \frac{1}{\sqrt{X^2}},
\ee
we cancel the divergence contained in it with the local term. As a result, the retarded field reads as
\be
\label{eq:5D_RF_massless}
\Psi(x;y) = - \frac{g}{4\pi^2} \int_{-\infty}^{\hat{\tau}} d\tau \, \frac{(aX)-1}{(vX)^2 \sqrt{X^2}}.
\ee

Calculating the field derivative and eliminating from it the local term arising from the differentiation of the upper integration limit by use of the similar integration by parts with the Eq. \eqref{eq:5D_int_parts_rel}, we obtain
\be
\partial_M \Psi(x;y) = - \frac{g}{4\pi^2} \int_{-\infty}^{\hat{\tau}} d\tau \left \lbrack \frac{(\dot{a}X)}{(vX)^3 \sqrt{X^2}} X_M - 3 \frac{((aX)-1)^2}{(vX)^4 \sqrt{X^2}} X_M - 3 \frac{(aX)-1}{(vX)^3 \sqrt{X^2}} v_M + \frac{a_M}{(vX)^2 \sqrt{X^2}} \right \rbrack.
\ee
In accordance with the Rohrlich-Teitelboim approach, extracting the leading $\hat{\rho}$-asymptotics of the field derivative by use of the relation
\be
X^M = \hat{\rho} \hat{c}^M + Z^M, \quad Z^M = \hat{z}^M - z^M,
\ee
we find the radiated part of five-dimensional massless scalar field as
\be
\label{eq:5D_massless_rad}
\lbrack \partial_M \Psi \rbrack^{\rm rad} = - \frac{g \hat{c}_M}{2^{5/2} \pi^2 \hat{\rho}^{3/2}} \int_{-\infty}^{\hat{\tau}} d\tau \left \lbrack \frac{(\dot{a}\hat{c})}{(v\hat{c})^3 \sqrt{(Z\hat{c})}} - 3 \frac{(a\hat{c})^2}{(v\hat{c})^4 \sqrt{(Z\hat{c})}} \right \rbrack.
\ee
Note that, due to the Huygens principle violation in odd dimensions, the radiated part of five-dimensional massless field \eqref{eq:5D_massless_rad} depends on the entire history of the charge's motion preceding the retarded time $\hat{\tau}$, rather that on its state at this moment, in contrast with the four-dimensional massless theory \eqref{eq:4D_massless_rad}.

\subsection{Emitted part of 4D massive field}

We extract the leading $\hat{\rho}$-asymptotics of the derivative of four-dimensional massive scalar field and demonstrate that it satisfies Eqs. \eqref{eq:CC_dim_red} and \eqref{eq:massless_limit_cond} and corresponds, thus, to the radiated part of the four-dimensional massive field derivative.

Substituting the scalar current \eqref{eq:source_def} into Eq. \eqref{eq:RF_mass_decomp_3} and using Eq. \eqref{eq:d-func_comp_arg} we arrive at the retarded four-dimensional massive scalar field in form
\be
\psi(x|\mu) = - \frac{g}{4\pi} \int_{-\infty}^{\hat{\tau}} d\tau \left \lbrack \frac{\mu J_1 (\mu \sqrt{X^2})}{\sqrt{X^2}} - \frac{\delta(\tau - \hat{\tau})}{\hat{\rho}} \right \rbrack,
\ee
where we used relation $J_0(0)=1$ \cite{zwillinger2014table}. By analogy with a five-dimensional massless field, using the relation
\be
\label{eq:int_parts_rel}
\frac{\mu J_1 (\mu \sqrt{X^2})}{\sqrt{X^2}} = \frac{1}{(vX)} \frac{d}{d\tau} J_0 (\mu \sqrt{X^2}),
\ee
we eliminate the local term integrating by parts, getting
\be
\label{eq:4D_RF_mass}
\psi(x|\mu) = - \frac{g}{4\pi} \int_{-\infty}^{\hat{\tau}} d\tau \, \frac{(aX)-1}{(vX)^2} J_0(\mu \sqrt{X^2}).
\ee

Let us show that the obtained retarded four-dimensional massive field satisfies Eq. \eqref{eq:dim_red}. Using the relation for the Bessel function \cite{zwillinger2014table}
\be
\label{eq:Bessel_int}
\int_{0}^{\infty} dx \, J_0(bx) = \frac{1}{b},
\ee
we calculate the integral over the Kaluza-Klein masses arriving at the equation
\be
\frac{1}{\pi} \int_{0}^{\infty} d\mu \, \psi(x|\mu) = - \frac{g}{4\pi^2} \int_{-\infty}^{\hat{\tau}} d\tau \, \frac{(aX)-1}{(vX)^2 \sqrt{X^2}},
\ee
which coincides with the retarded five-dimensional massless field \eqref{eq:5D_RF_massless}. Let us, also, check that in the massless limit the Eq. \eqref{eq:4D_RF_mass} matches the Eq. \eqref{eq:4D_RF_massless}. To show it, we integrate by parts
\be
\label{eq:Massless_lim_comput}
\psi(x|0) = - \frac{g}{4\pi} \int_{-\infty}^{\hat{\tau}} d\tau \, \frac{(aX)-1}{(vX)^2} = \frac{g}{4\pi} \int_{-\infty}^{\hat{\tau}} d\tau \, \frac{d}{d\tau} \frac{1}{(vX)} = \frac{g}{4\pi} \left. \frac{1}{(vX)} \right \vert_{-\infty}^{\hat{\tau}} = \frac{g}{4\pi\hat{\rho}}.
\ee
Here we assumed that the charge moves inside the compact region of space.

Computing derivative of the retarded field \eqref{eq:4D_RF_mass} and eliminating from it the local term resulting from the differentiation of the upper integration limit by use of the integration by parts with account for Eq. \eqref{eq:int_parts_rel}, we find the field derivative as
\be
\label{eq:4D_RF_mass_deriv}
\partial_\mu \psi(x|\mu) = - \frac{g}{4\pi} \int_{-\infty}^{\hat{\tau}} d\tau \left \lbrack \frac{(\dot{a}X)}{(vX)^3} X_\mu - 3 \frac{((aX)-1)^2}{(vX)^4} X_\mu - 3 \frac{(aX)-1}{(vX)^3} v_\mu + \frac{a_\mu}{(vX)^2} \right \rbrack J_0(\mu \sqrt{X^2}).
\ee
Here, all the terms of the integrand contain the same damping factor $J_0(\mu \sqrt{X^2})$, which determines the contribution of the history of the charge's motion into the field. Therefore, it gives more control on the extraction of the leading $\hat{\rho}$-asymptotics of the field derivative.

Extracting the leading $\hat{\rho}$-asymptotics of the field derivative we consider the quantity $\mu \hat{\rho}$ in the argument of the Bessel function to have a finite value, since the spectral functions in the five-dimensional massless theory \eqref{eq:dim_red} and in DGP model \eqref{eq:4+1_RGF_KKI_1} have a finite value at the mass range $\mu \sim 0$ close to zero. In this range, ratio of the Lorentz-invariant distance $\hat{\rho}$ to the characteristic length scale defined by the Kaluza-Klein mass $1/\mu$ is finite. As a result, it is this ratio that determines the intensity of radiation leakage into the bulk with the distance from the source. Thus, extracting the leading $\hat{\rho}$-asymptotics of Eq. \eqref{eq:4D_RF_mass_deriv} we get
\be
\label{eq:4D_RF_rad_deriv}
\lbrack \partial_\mu \psi(x|\mu) \rbrack^{\rm rad} = - \frac{g \hat{c}_\mu}{4\pi\hat{\rho}} \int_{-\infty}^{\hat{\tau}} d\tau \left \lbrack \frac{(\dot{a}\hat{c})}{(v\hat{c})^3} - 3 \frac{(a\hat{c})^2}{(v\hat{c})^4} \right \rbrack J_0(\mu \sqrt{2\hat{\rho} (Z\hat{c})}).
\ee
Let us verify that the obtained leading $\hat{\rho}$-asymptotics satisfies Eq. \eqref{eq:CC_dim_red} and \eqref{eq:massless_limit_cond}. The Eq. \eqref{eq:CC_dim_red} is easily proved by use of the Eqs.\eqref{eq:Bessel_int} and (\ref{eq:RCQ_dim_rel_1}--\ref{eq:RCQ_dim_rel_2})
\be
\frac{1}{\pi} \int_0^{\infty} d\mu \, \lbrack \partial_\mu \psi(x|\mu) \rbrack^{\rm rad} = - \frac{g \hat{c}_\mu}{2^{5/2} \pi^2 \hat{\rho}^{3/2}} \int_{-\infty}^{\hat{\tau}} d\tau \left \lbrack \frac{(\dot{a} \hat{c})}{(v \hat{c})^3 \sqrt{(Z \hat{c})}} - 3 \frac{(a \hat{c})^2}{(v \hat{c})^4 \sqrt{(Z \hat{c})}} \right \rbrack,
\ee
where the obtained integral coincides with the emitted part of five-dimensional massless field on the brane \eqref{eq:5D_massless_rad}. The correctness of the massless limit is proved by analogy with Eq.\eqref{eq:Massless_lim_comput}
\begin{multline}
\lbrack \partial_\mu \psi(x|0) \rbrack^{\rm rad} = - \frac{g \hat{c}_\mu}{4\pi\hat{\rho}} \int_{-\infty}^{\hat{\tau}} d\tau \left \lbrack \frac{(\dot{a}\hat{c})}{(v\hat{c})^3} - 3 \frac{(a\hat{c})^2}{(v\hat{c})^4} \right \rbrack = - \frac{g \hat{c}_\mu}{4\pi\hat{\rho}} \int_{-\infty}^{\hat{\tau}} d\tau \frac{d}{d\tau} \frac{(a \hat{c})}{(v \hat{c})^3} = - \frac{g \hat{c}_\mu}{4\pi\hat{\rho}} \left. \frac{d}{d\tau} \frac{(a \hat{c})}{(v \hat{c})^3} \right \vert_{-\infty}^{\hat{\tau}} = \\ = - \frac{g (\hat{a}\hat{c}) \hat{c}_\mu}{4 \pi \hat{\rho}},
\end{multline}
where the resulting equation matches Eq. \eqref{eq:4D_massless_rad}. Here, to obtain the correct massless limit, an additional asymptotic condition was imposed on the charge world line similar to that found in Ref. \cite{Teitelboim1970} and corresponding to the uniform rectilinear motion of the charge in the distant past
\be
a^\mu \to 0, \quad \tau \to - \infty.
\ee
However, in practice, the world line of the charge in many cases can be approximated by simpler laws of motion, for which this condition does not holds, due to the presence in the integrand in Eq. \eqref{eq:4D_RF_rad_deriv} of the damping factor $J_0(\mu \sqrt{2\hat{\rho} (Z\hat{c})})$ making the world lines effectively equivalent.

Thus, it was demonstrated that the emitted part of derivative of the four-dimensional massive scalar field can be extracted, by analogy with the massless fields, as its leading $\hat{\rho}$-asymptotics, in accordance with the Rohrlich-Teitelboim approach.

\section{Leakage of radiation from the brane}\label{V}

We now turn to the calculation of radiation produced by the point charge in the scalar DGP model. In accordance with Eqs. \eqref{eq:4+1_DGP_rad_gen} and \eqref{eq:4D_RF_rad_deriv}, the emitted part of derivative of the field on the brane takes the form
\be
\label{eq:4+1_DGP_rad_h+m_ints}
\lbrack \partial_\mu \varphi(x;0) \rbrack^{\rm rad} = \frac{g \hat{c}_\mu}{8 \pi^2 M_{5}^{3} \hat{\rho}} \int_{0}^{\infty} d\mu \, \rho(\mu) \int_{-\infty}^{\hat{\tau}} d\tau \left \lbrack \frac{(\dot{a}\hat{c})}{(v\hat{c})^3} - 3 \frac{(a\hat{c})^2}{(v\hat{c})^4} \right \rbrack J_0(\mu \sqrt{2\hat{\rho} (Z\hat{c})}),
\ee
where the spectral function is given by Eq. \eqref{eq:4+1_RGF_KKI_1}. Note that, in accordance with the Huygens principle violation in the five-dimensional bulk of DGP model, radiated part of the field derivative on the brane \eqref{eq:4+1_DGP_rad_h+m_ints} depends on the entire history of the source motion preceding the retarded time $\hat{\tau}$, by analogy with the massless five-dimensional field \eqref{eq:5D_massless_rad}.

In the Eq. \eqref{eq:4+1_DGP_rad_h+m_ints}, one can first calculate the integral over the Kaluza-Klein masses to get an explicit form of the damping factor in the integral over the history of charge motion. However, it is convenient to first calculate the integral over the history of motion, and only after that the integral over the Kaluza-Klein masses.

\subsection{Radiation from the non-relativistic charge}

In what follows, we will only be interested in the radiation of a non-relativistic charge. In this case, equation for the emitted part of derivative of the field on the brane is significantly simplified.

To calculate the non-relativistic approximation of the Eq. \eqref{eq:4+1_DGP_rad_h+m_ints} we assume that:
\begin{itemize}

\item
the charge is non-relativistic $|\mathbf{v}| \ll 1, \; \forall \tau$;

\item
the charge moves inside a compact region of space $|\mathbf{z}| \leq d, \; \forall \tau$ ($d$ is the characteristic size of the region);

\item
the observation point is far from this region $d \ll r$.

\end{itemize}
Thus, covariant retarded quantities are expanded in small parameters $\mathbf{v}$ and $|\mathbf{z}|/r$ up to first order as follows
\begin{align}
&\hat{\tau} \simeq \bar{t} + \mathbf{n}\bar{\mathbf{z}}, \quad \mathbf{n} = \mathbf{x}/r, \\
&\hat{\rho} \simeq r \left( 1 - \mathbf{n}\bar{\mathbf{v}} - \frac{1}{r} \mathbf{n}\bar{\mathbf{z}} \right), \\
&\hat{c}^\mu \simeq \left \lbrack 1 + \mathbf{n}\bar{\mathbf{v}}; \, \mathbf{n} \left( 1 + \mathbf{n}\bar{\mathbf{v}} + \frac{1}{r} \mathbf{n}\bar{\mathbf{z}} \right) - \frac{1}{r}\bar{\mathbf{z}} \right \rbrack,
\end{align}
where $\bar{t} = t - r$ is the retarded proper time computed up to the leading contribution, and all the barred quantities correspond to this moment. We, also, introduce an additional space-like vector with the same order of magnitude as the charge velocity \cite{Khlopunov:2022ubp}
\be
\mathbf{s}(t) = \frac{\bar{\mathbf{z}} - \mathbf{z}(t)}{\bar{t} - t}, \quad |\mathbf{s}| \sim |\mathbf{v}|.
\ee
Here we replaced the charge proper time with the coordinate time $\tau = t$, given their equivalence in the non-relativistic approximation.

Thus, the combinations of covariant retarded quantities involved in Eq. \eqref{eq:4+1_DGP_rad_h+m_ints} are expanded up to the first order in small parameters as follows
\begin{align}
\label{eq:nr_exp_1}
& (\dot{a}\hat{c}) \simeq - \dot{\mathbf{a}} \mathbf{n}, \quad (a\hat{c}) \simeq - \mathbf{a} \mathbf{n}, \\
\label{eq:nr_exp_2}
& (v\hat{c}) \simeq 1 + \mathbf{n}(\bar{\mathbf{v}} - \mathbf{v}), \\
\label{eq:nr_exp_3}
& \sqrt{\hat{\rho} (Z\hat{c})} \simeq \sqrt{r(\bar{t} - t)} \left \lbrack 1 - \frac{1}{2} \frac{\mathbf{n}\bar{\mathbf{z}}}{r} - \frac{1}{2} \mathbf{n} \mathbf{s} \right \rbrack,
\end{align}
given that in order for the charge to be non-relativistic throughout the history of motion, its acceleration and its derivatives must, also, be small quantities. As a result, by use of the Eqs. (\ref{eq:nr_exp_1}--\ref{eq:nr_exp_3}) we obtain the non-relativistic approximation for the emitted part of the field derivative on the brane
\be
\label{eq:4+1_DGP_rad_non-rel}
\lbrack \partial_\mu \varphi(x;0) \rbrack^{\rm rad} \simeq - \frac{g \bar{c}_\mu}{8\pi^2 M_{5}^{3} r} \int_{0}^{\infty} d\mu \, \rho(\mu) \int_{-\infty}^{\bar{t}} dt' \, \dot{\mathbf{a}}\mathbf{n} \, J_0(\mu \sqrt{2r(\bar{t} - t')}), \quad \bar{c}_\mu = \lbrack 1; - \mathbf{n} \rbrack.
\ee
Note that in the scalar DGP model, up to the leading order in non-relativistic expansion, charge moving with constant acceleration does not radiate, in contrast to the four-dimensional massless theory \eqref{eq:4D_massless_rad}. Analogous result was obtained in the theory of five-dimensional massless scalar field \cite{Khlopunov:2022ubp}. However, in Ref. \cite{Khlopunov:2022ubp} it was, also, demonstrated that in the five-dimensional General Relativity a point mass with constant acceleration generates gravitational radiation. Based on this, we assume that in the gravitational DGP model a uniformly accelerated mass also radiates gravitational waves.

\subsection{Radiation from the charge on a circular orbit}

Let us calculate the effective four-dimensional energy flux of scalar radiation through a 2-sphere on the brane -- the observed effective power of scalar radiation into the brane -- for the case of a non-relativistic charge moving along a circular trajectory. The resulting equation will characterize the intensity of leakage of scalar radiation from the brane into the extra dimension.

The world line of a non-relativistic charge on a circular orbit on the brane reads as
\be
\mathbf{z}(t) = \lbrace R_0 \cos{\omega_0 t}, R_0 \sin{\omega_0 t}, 0 \rbrace,
\ee
where $R_0$ is the radius of the orbit and $\omega_0$ is the frequency of orbital motion. Substituting this world line in Eq. \eqref{eq:4+1_DGP_rad_non-rel}, for the integral over the history of motion we find \cite{zwillinger2014table}
\be
\int_{-\infty}^{\bar{t}} dt' \, \dot{\mathbf{a}}\mathbf{n} \, J_0(\mu \sqrt{2r(\bar{t} - t)}) = R_0 \omega_0^2 \sin\theta \left \lbrack \sin(\omega_0\bar{t} - \phi) \sin{\frac{r\mu^2}{2\omega_0}} - \cos(\omega_0\bar{t} - \phi) \cos{\frac{r\mu^2}{2\omega_0}} \right \rbrack,
\ee
where we use the spherical coordinates on the brane $\mathbf{n} = \lbrace \cos \phi \sin \theta, \sin \phi \sin \theta, \cos \theta \rbrace$. Calculating the integral over the Kaluza-Klein masses \cite{zwillinger2014table}, we arrive at the radiated part of field derivative on the brane in form
\begin{multline}
\label{eq:4+1_DGP_rad_circ}
\lbrack \partial_\mu \varphi(x;0) \rbrack^{\rm rad} = \frac{g \bar{c}_\mu}{8\pi M_{4}^{2} r} R_0 \omega_0^2 \sin\theta \left \lbrack \cos{(\omega_0\bar{t} - \phi - x^2)} - \sqrt{2} \cos{\left( \omega_0\bar{t} - \phi - x^2 - \frac{\pi}{4} \right)} C(x) + \right. \\ \left. + \sqrt{2} \sin{\left( \omega_0\bar{t} - \phi - x^2 - \frac{\pi}{4} \right)} S(x) \right \rbrack, \quad x \equiv \sqrt{\frac{rM_{c}^{2}}{2\omega_0}},
\end{multline}
where $C(x)$ and $S(x)$ are the cosine and sine Fresnel integrals, respectively.

Substituting Eq. \eqref{eq:4+1_DGP_rad_circ} into the Eq. \eqref{eq:4D_eff_EMT}, we find the radiated part of the effective four-dimensional energy-momentum tensor of the field on the brane
\begin{multline}
\label{eq:EMT_rad_circ}
\Theta_{\mu\nu}^{\rm rad} = \frac{g^2 \bar{c}_\mu \bar{c}_\nu}{32 \pi^2 M_{4}^{2} r^2} R_0^2 \omega_0^4 \sin^2 \theta \left \lbrack \cos{(\omega_0\bar{t} - \phi - x^2)} - \sqrt{2} \cos{\left(\omega_0\bar{t} - \phi - x^2 - \frac{\pi}{4}\right)} C(x) + \right. \\ \left. + \sqrt{2} \sin{\left(\omega_0\bar{t} - \phi - x^2 - \frac{\pi}{4}\right)} S(x) \right \rbrack^2.
\end{multline}
Substituting the obtained radiated part of the effective energy-momentum tensor in Eq. \eqref{eq:brane_rad_power}, after integration over the angular variables \cite{zwillinger2014table} we arrive at the effective four-dimensional energy flux of scalar radiation through a distant 2-sphere of radius $r$ on the brane -- the effective four-dimensional power of scalar radiation of the charge into the brane
\be
\label{eq:DGP_brane_rad_circ}
W_{\rm br}^{\rm circ}(r) = \frac{g^2 R_0^2 \omega_0^4}{24 \pi M_{4}^{2}} \Big \lbrack 1 - 2C(x) - 2S(x) + 2C^2(x) + 2S^2(x) \Big \rbrack.
\ee
First of all, we note that in the limit $M_5 \to 0$, which transforms the model under consideration into the ordinary four-dimensional theory of a massless scalar field
\be
M_{5}^{3} \, {^{(5)}}\square \varphi + M_{4}^{2} \, \delta(y) {^{(4)}}\square \varphi = - \frac{1}{2} j(x) \delta(y) \, \xrightarrow{M_5 \to 0} \, {^{(4)}}\square \varphi = - \frac{1}{2M_{4}^{2}} j(x),
\ee
the effective radiation power, also, transforms into the standard four-dimensional form, losing its dependence on the radius of the sphere through which it passes
\be
\lim_{x \to 0} C(x), S(x) = 0 \quad \Longrightarrow \quad W_{\rm br}^{\rm circ}(r) \xrightarrow{M_5 \to 0} \frac{g^2 R_0^2 \omega_0^4}{24 \pi M_{4}^{2}} = {\rm Const}.
\ee

\begin{figure}[t]
\center{\includegraphics[width=0.6\linewidth]{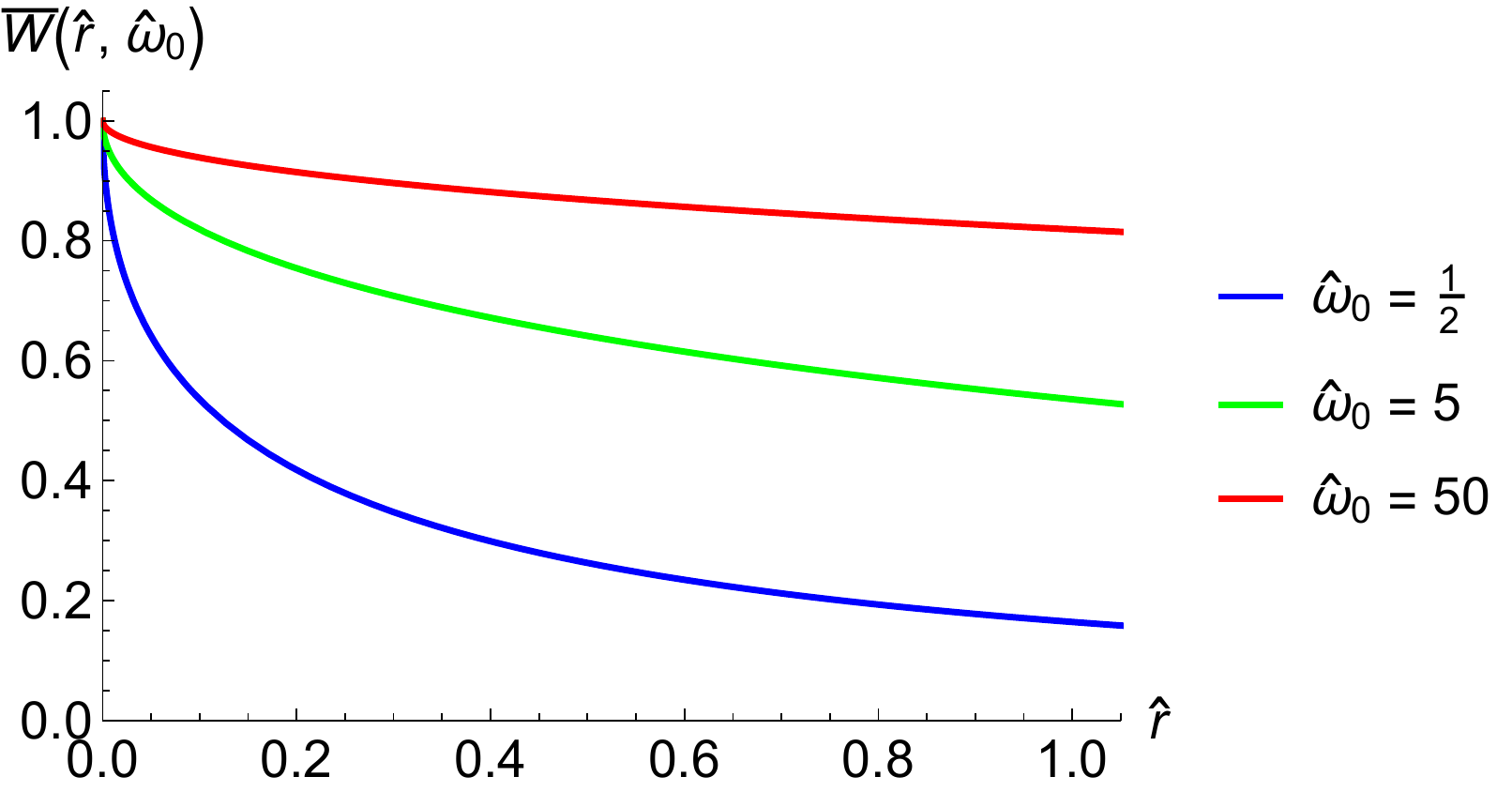}}
\caption{Dependence of the effective four-dimensional radiation energy flux through a 2-sphere on the brane on its radius for various frequencies of the signal.}
\label{fig:4}
\end{figure}

Let us analyze the properties of the obtained effective four-dimensional power of scalar radiation into the brane. For convenience, we construct the normalized energy flux
\be
\overline{W}_{\rm br}^{\rm circ}(r) = \frac{W_{\rm br}^{\rm circ}(r)}{W_{\rm br}^{\rm circ}(0)} = 1 - 2C(x) - 2S(x) + 2C^2(x) + 2S^2(x).
\ee
We, also, introduce the dimensionless parameters of distance and frequency
\be
\hat{r} = rM_c, \quad \hat{\omega}_0 = \frac{\omega_0}{M_c}.
\ee
As a result, the normalized effective power of radiation into the brane takes the form
\be
\label{eq:norm_en_flux}
\overline{W}_{\rm br}^{\rm circ}(\hat{r},\hat{\omega}_0) = 1 - 2C(\sqrt{\hat{r}/2\hat{\omega}_0}) - 2S(\sqrt{\hat{r}/2\hat{\omega}_0}) + 2C^2(\sqrt{\hat{r}/2\hat{\omega}_0}) + 2S^2(\sqrt{\hat{r}/2\hat{\omega}_0}).
\ee
From Fig. \eqref{fig:4} we find that the effective energy flux monotonically decreases with the radius of the sphere through which it passes, in accordance with the gradual leakage into the bulk of the field perturbations propagating along the brane (a metastable scalar particle decays on the brane \cite{Gabadadze:2003ck,Gabadadze:2004dq}). Note, also, that the higher the frequency of orbital motion, and hence the frequency of the signal, the lower is the intensity of leakage of scalar radiation into the bulk with the distance from its source (see Fig. \eqref{fig:4}), in accordance with the infrared transparency of the bulk in DGP model \cite{Dvali:2000xg,Brown:2016gwv}. In particular, in agreement with the brane transmission coefficient \eqref{eq:Trans_coef} with the maximum at zero frequency and decreasing monotonically with the signal frequency, high-frequency signals turn out to be quasi-localized on the brane.

Based on the obtained effective four-dimensional power of scalar radiation into the brane \eqref{eq:norm_en_flux}, we can preliminarily analyze the possibility of experimental observation of the effect of gravitational wave leakage into the extra dimension. We choose the parameters of the DGP model in such a way that the crossover scale corresponds to the modern Hubble radius $r_c \sim H_0^{-1} \sim 10^{26} \, {\rm m}$, or $M_c \sim 10^{-42} \, {\rm GeV}$ \cite{Deffayet:2001pu,Deffayet:2002sp}. We find that for signals with frequencies in the sensitivity range of current and future gravitational wave observatories \cite{Moore:2014lga}, the intensity of radiation leakage into the bulk is extremely small even at cosmological distances $r \sim r_c$ from the signal source:
\begin{itemize}

\item
LIGO observatory: $\nu_{\rm ch} \sim 10^2 \, {\rm Hz} \; \Longrightarrow \; \hat{\omega}_0 \sim 10^{20} \; \Longrightarrow \; \overline{W}(\hat{r}=1,\,\hat{\omega}_0=10^{20}) \simeq 0.9999999998$;

\item
LISA observatory: $\nu_{\rm ch} \sim 10^{-3} \, {\rm Hz} \; \Longrightarrow \; \hat{\omega}_0 \sim 10^{15} \; \Longrightarrow \; \overline{W}(\hat{r}=1;\,\hat{\omega}_0=10^{15}) \simeq 0.99999994$.
\end{itemize}

Therefore, experimental observation of the effect of gravitational wave leakage into the extra dimension by use of the modern and future gravitational wave observatories seems to be extremely unlikely. In turn, signals whose leakage hypothetically could be detected on the scales $r \sim r_c$, i.e., for which
\be
\overline{W}(\hat{r}=1;\,\hat{\omega}_0) \lesssim 0.99,
\ee
should have frequencies of order $\omega_0 \lesssim 10^{-13} \, {\rm Hz}$, which is much lower than the sensitivity ranges of gravitational-wave observatories \cite{Moore:2014lga}. On the other hand, in order for the leakage of signals from the sensitivity ranges of gravitational-wave observatories to be detectable, the crossover radius must have a much smaller value. In particular, in case of the LISA observatory, which will detect the lowest frequency gravitational-wave signals, the crossover radius should be $r_c \sim 10^{17} \, {\rm m}$.

\section{Conclusions}\label{VI}

The goal of this work was to explore the leakage of radiation into extra dimensions, corresponding to the metastable nature of effective graviton on the brane in DGP model, within a simple five-dimensional  scalar field analog of the DGP model. As a source of the field, we considered a point charge moving along a fixed world line on the brane. Presumably, this model correctly reproduces features of the gravitational DGP model, thus providing the estimate of the effect for true gravitational waves. We clarified a number of technical details in such a construction and revealed features associated with violation of the Huygens principle in the bulk.   The main features of the field behaviour in the bulk and on the brane were also illustrated using lower dimensional DGP-like models.

As field on the brane is composed of continuous spectrum of massive Kaluza-Klein modes, to extract its emitted part we generalised the Rohrlich-Teitelboim approach to radiation to the case of four-dimensional massive field using its connections with massless fields in dimensions four and five. In particular, we demonstrated that, due to the Huygens principle violation in the five-dimensional bulk, emitted part of the field on the brane depends on the history of charge's motion, preceeding the retarded time, rather than on its state at this moment, in contrast to the four-dimensional massless theory. We obtained equation for the effective four-dimensional energy flux of scalar radiation through the 2-sphere on the brane produced by the non-relativistic charge on a circular orbit \eqref{eq:DGP_brane_rad_circ}. It was found that this effective power of radiation into the brane monotonically decreases with the radius of the sphere, through which it passes (see Fig. \eqref{fig:4}), in accordance with the metastable nature of corresponding scalar particle on the brane resulting into the leakage of radiation to the extra dimension. Also, it was shown that, in accordance with the infrared transparency of the bulk in the DGP model, radiation leakage intensity is higher for the low frequency signals (see Fig. \eqref{fig:4}).

Using the obtained effective power of radiation into the brane, we estimated the radiation leakage intensity and analysed the possibility of experimental observation of this effect. For the ``realistic'' choice of the crossover scale $r_c \sim 10^{26}$ m, we demonstrated that within the framework of our model effect of leakage of radiation into the extra dimension is negligible even for the signals from the frequency range accessible for the LISA observatory. This effect can be significant either in the case of extremely low frequency signals, which are unavailable for the experimental observation by current and future gravitational-wave observatories, either in the case of small values of crossover scale, which make the gravitational DGP model phenomenologically non-viable.

Analogous effect was observed in Ref. \cite{Gregory:2000jc} within the framework of modified Randall-Sundrum model, which also contains the metastable effective graviton on the brane. In our work an explicit expression for the effective power of radiation into the brane is obtained, providing the estimate of the intensity of radiation leakage into the extra dimension.

\section*{Acknowledgements}

The work of M. Kh. was supported by the “BASIS” Foundation Grant No. 20-2-10-8-1. This research was also supported by the Russian Foundation for Basic Research on the project 20-52-18012Bulg-a, and the Scientific and Educational School of Moscow State University “Fundamental and Applied Space Research”.

\appendix

\section{Field of the static charge on the brane}

To calculate the field of static charge on the brane we start with Eq. \eqref{eq:ret_field_br} and Eqs. (\ref{eq:4+1_RGF_KKI_1}--\ref{eq:4+1_RGF_KKI_2}). In the case of charge resting on the brane at the origin $z^\mu(\tau) = \lbrack \tau, \mathbf{0} \rbrack$, scalar current \eqref{eq:source_def} takes simple form
\be
\label{eq:sc_cur_static}
j(x) = g \delta^{(3)}(\mathbf{x}).
\ee

Substituting it into the Eq. \eqref{eq:ret_field_br}, for the integral with four-dimensional massive Green's function we find \cite{zwillinger2014table}
\be
\label{eq:KK_Yukawa_pot}
\int d^4x' \, G_{\rm 4D}(x-x'|\mu) \, j(x') = \frac{g e^{-\mu r}}{4 \pi r}, \quad r = |\mathbf{x}|.
\ee
Non-surprisingly, Eq. \eqref{eq:KK_Yukawa_pot} corresponds to the Yukawa potential of static charge, interacting with four-dimensional massive field.

As a result, the field on the brane is given by the integral over continuous Kaluza-Klein spectrum of four-dimensional Yukawa potentials
\be
\varphi(x;0) = - \frac{g M_c^2}{8 \pi^2 M_5^3 r} \int_0^\infty d\mu \frac{e^{-\mu r}}{\mu^2 + M_c^2}.
\ee
This integral is easily calculated \cite{zwillinger2014table} resulting into the field of static charge on the brane in form
\be
\label{eq:field_stat_br}
\varphi(x;0) = - \frac{g}{4 \pi^2 M_4^2 r} \left \lbrack {\rm Ci}(M_c r) \sin{M_c r} + \frac{1}{2} \left( \pi - 2 \, {\rm Si}(M_c r) \right) \cos{M_c r} \right \rbrack,
\ee
where ${\rm Si}(x)$ and ${\rm Ci}(x)$ are sine and cosine integrals, correspondingly. Eq. \eqref{eq:field_stat_br} coincides with the field of static charge on the brane obtained within the normal branch of DGP model \cite{Dvali:2000hr}. Thus, the radiation boundary condition in the bulk corresponds to the choice of normal branch of DGP model.


\end{document}